\documentclass[pdflatex,sn-mathphys-num, Numbered]{sn-jnl}% Math and Physical Sciences Numbered Reference Style
\usepackage{graphicx}
\usepackage{multirow}
\usepackage{amsmath}
\usepackage{amssymb}
\usepackage{amsfonts}
\usepackage{amsthm}
\usepackage{mathrsfs}
\usepackage[title]{appendix}
\usepackage{xcolor}
\usepackage{textcomp}
\usepackage{manyfoot}
\usepackage{booktabs}
\usepackage{algorithm}
\usepackage{algorithmicx}
\usepackage{algpseudocode}
\usepackage{listings}
\usepackage{multirow}

\theoremstyle{thmstyleone}%
\theoremstyle{thmstyletwo}%

\theoremstyle{thmstylethree}%
\newtheorem{definition}{Definition}%

\newcommand{\Cov}{\mathrm{Cov}}

\begin{document}

\title[Article Title]{Wavelet Localisation and Local Modulation Freezing in MRW Unwrapping
%Article Title
}


\author[1]{\fnm{Mateusz} \sur{Polakowski}}\email{mm.polakowsk@student.uw.edu.pl}

\author*[1, 2]{\fnm{Zbigniew R.} \sur{Struzik}}\email{z.r.struzik@p.u-tokyo.ac.jp}

\affil[1]{\orgdiv{Faculty of Physics}, \orgname{University of Warsaw}, \orgaddress{\street{Pasteura 5}, \postcode{02-093}, \city{Warsaw}, \country{Poland}}}

\affil[2]{\orgdiv{Graduate School of Education}, \orgname{University of Tokyo}, \orgaddress{\street{7‑3‑1 Hongo, Bunkyo‑ku}, \city{Tokyo}, \postcode{113‑0033}, \country{Japan}}}


\abstract{We develop a localised wavelet formulation of multifractal random walk (MRW) unwrapping based on the concept of local multiplicative modulation freezing. The framework is motivated by the observation that finite-support wavelet localisation may induce approximate local factorisation of multiplicatively modulated stochastic fields, allowing the modulation component to become effectively frozen within sufficiently localised probing domains. Within this regime, logarithmic wavelet amplitudes admit an approximate additive decomposition linking local wavelet statistics directly to the underlying modulation field.

This viewpoint reformulates covariance-based MRW unwrapping as a localised multiscale operator problem in which wavelet coefficients act as finite-support probes of multiplicative organisation. Unlike classical global covariance approaches, the validity of the approximation depends explicitly on support geometry, scale-dependent overlap, and residual multiscale mixing generated by internal modulation variability. We show that these effects naturally produce finite-scale deviations from ideal logarithmic covariance scaling and lead to structured covariance distortions whose form depends on the interaction between the modulation field and the geometry of the wavelet representation.

The resulting framework provides a localised interpretation of multiplicative modulation extraction that differs conceptually from traditional multifractal wavelet approaches based primarily on global statistical characterisation. In particular, localisation itself becomes the operational mechanism enabling multiscale probing of local stochastic organisation. Numerical investigations using orthonormal wavelet decompositions support the proposed interpretation and demonstrate the emergence of scale-dependent freezing regimes, residual covariance mixing, and finite-support breakdown effects consistent with the theoretical formulation.

Beyond the immediate context of MRW unwrapping, the proposed framework suggests a broader connection between wavelet localisation, local regularity organisation, and finite-support multiscale stochastic operators. From this perspective, wavelet localisation becomes not merely a decomposition principle, but an operational mechanism for probing localised multiscale structure.
}




\maketitle

\section{Introduction}\label{Intro}
Intermittent multiscale organisation is a defining property of many complex physical and stochastic systems. Examples include fully developed turbulence, geophysical transport, financial fluctuations, stochastic textures, biological variability, and heterogeneous image structure~\cite{Kang14,Muzy06,Schmitt99,SerGiacomi15}. Such systems are characterised by highly nonuniform fluctuation organisation in which relatively smooth regions coexist with localised bursts of intense activity across a broad range of scales. Classical Gaussian models fail to reproduce this behaviour because they do not account for the strong scale-dependent variability of local fluctuation amplitudes~\cite{Jimenez2000}. Understanding the statistical and geometrical structure of intermittency therefore remains a central problem in multiscale physics.

One of the most influential approaches to intermittent multiscale processes is based on multiplicative stochastic organisation~\cite{Borlandetal2005}. In this viewpoint, observed fluctuations are modeled as rapidly varying carrier processes modulated by slowly varying stochastic amplitude fields. This general framework underlies multiplicative cascades~\cite{Mandelbrot1974}, multifractal random walks (MRW)~\cite{Bacryetal2000}, Gaussian multiplicative chaos models~\cite{RobertVargas2010}, and a broad class of multifractal stochastic processes~\cite{RhodesVargas2014, Abryetal2009}. In particular, the multifractal random walk framework introduced by Bacry, Muzy, and collaborators provided a statistically tractable continuous formulation of intermittency based on logarithmically correlated stochastic modulation fields~\cite{Abryetal2009, Borlandetal2005, Bacryetal2000}. Such models successfully reproduce nontrivial scaling laws, heavy-tailed fluctuations, and multiscale covariance structure observed in a wide range of complex systems.
A central consequence of multiplicative organisation is the emergence of logarithmic covariance relations~\cite{RhodesVargas2014}. In MRW-type models, the modulation field typically satisfies
\begin{equation}
\mathrm{Cov}[\omega(x),\omega(x+r)] \sim -\lambda^2\log r,
\end{equation}
over an appropriate scaling range, where $\lambda^2$ controls the strength of intermittency. These covariance laws provide one of the most robust statistical signatures of multifractal modulation and have motivated a variety of estimation and inverse formulations. In particular, Lakhal and collaborators demonstrated that covariance structure may be exploited operationally in order to estimate or “unwrap” hidden modulation fields from observed intermittent processes~\cite{Lakhal25}.

At the same time, wavelet methods have played a fundamental role in the development of multifractal analysis. The introduction of wavelet-based singularity analysis by Mallat and Hwang~\cite{MallatHwang1992}, followed by the wavelet transform modulus maxima (WTMM) framework developed extensively by Arneodo, Bacry, Muzy, and collaborators~\cite{Muzyetal1991, Muzyetal1993}, established wavelets as one of the principal tools for multiscale characterisation of singular and intermittent structures. Wavelet methods proved especially powerful because they simultaneously localise information in scale and position while remaining naturally connected to local regularity and singularity structure. Further developments by Jaffard and collaborators~\cite{Jaffard1997a, Jaffard1997b} established deep links between wavelet coefficients, H\"older regularity, multifractal spectra, and local scaling geometry.

Despite these advances, wavelet localisation has largely been used as an intermediate mechanism for estimating global statistical quantities such as multifractal spectra, partition functions, or scaling exponents~\cite{FrischParisi1985, Muzyetal1993, Wendtetal2007}. Even though locality constitutes the conceptual foundation of wavelet singularity analysis, the localised nature of wavelet coefficients themselves has rarely been interpreted operationally as a mechanism for local multiplicative probing. Existing multifractal formulations therefore remain predominantly global in character. Statistical characterisation is typically performed through ensemble averages, covariance functions, or global scaling laws rather than through explicitly localised multiscale operators~\cite{Lashermesetal2004, Wendtetal2007}.

The present work develops a different perspective. Instead of treating wavelets merely as multiscale representation tools, we interpret wavelet localisation itself as an operational mechanism capable of inducing approximate local multiplicative separation. More precisely, we investigate whether sufficiently localised wavelets may create finite-support observation domains over which multiplicative modulation becomes approximately frozen relative to the support of the analyzing wavelet.
This idea leads naturally to a localised multiscale reformulation of MRW unwrapping.

The central observation developed throughout the manuscript is that wavelet localisation induces an approximate local factorisation of the form
\begin{equation}
W_X(a,b) \approx e^{\omega(b)} W_\epsilon(a,b),
\end{equation}
or equivalently,
\begin{equation}
\log \lvert W_X(a,b) \rvert \approx \omega(b) + \log |W_\epsilon(a,b)|.
\end{equation} 

These relations constitute the basis of a localised modulation freezing approximation whose validity depends explicitly on wavelet support geometry, local regularity of the modulation field, and finite-scale multiscale mixing effects.

Importantly, the formulation developed here does not rely on a particular modulation ontology. Multiplicative cascades and MRW models are treated instead as important stochastic realisations of a broader class of multiplicatively modulated multiscale processes. The essential mechanism underlying the proposed framework is localisation itself: wavelets create finite-support probing domains whose geometry controls the validity of local multiplicative separation.
This perspective produces a conceptual shift from global statistical characterisation toward localised multiscale probing.

The manuscript develops this framework systematically. First, we introduce multiplicative modulation fields and review the covariance structure underlying MRW formulations. We then reinterpret wavelet transforms as localised multiscale probing operators whose finite-support geometry naturally induces local modulation freezing. The resulting approximation leads to a localised wavelet reformulation of MRW unwrapping together with an explicit interpretation of residual covariance distortions as signatures of finite-scale multiscale mixing. We further connect the freezing approximation to local regularity theory and wavelet-based singularity analysis in the sense of Jaffard.

The theoretical developments are supported through extensive numerical investigations using synthetic one-dimensional MRW realisations. Particular attention is devoted to support-induced covariance distortions, residual multiscale coupling, finite-scale breakdown of freezing validity, and scale-dependent reconstruction quality. The results demonstrate that many of the deviations observed numerically are not implementation artifacts but rather natural consequences of localised multiscale mixing generated by finite wavelet support.

Beyond its immediate implications for MRW unwrapping, the present framework suggests a broader interpretation of wavelet localisation in intermittent multiscale systems. Rather than functioning solely as representational tools, localised wavelet operators may provide direct access to local modulation structure itself. This viewpoint opens potential connections to localised multifractal analysis, nonstationary intermittency, inverse multiscale problems, stochastic texture modelling, and multiscale regularity theory in complex systems.

%The Introduction section, of referenced text~\cite{bib1} expands on the background of the work (some overlap with the Abstract is acceptable). The introduction should not include subheadings.

%Springer Nature does not impose a strict layout as standard however authors are advised to check the individual requirements for the journal they are planning to submit to as there may be journal-level preferences. When preparing your text please also be aware that some stylistic choices are not supported in full text XML (publication version), including coloured font. These will not be replicated in the typeset article if it is accepted. 

\section{Multiplicative Modulation and MRW Background}\label{sec2}
Intermittent multiscale fluctuations arise in a wide range of complex systems including turbulence, geophysical dynamics, finance, image textures, and stochastic transport processes~\cite{FrischParisi1985, SerGiacomi15, Schmitt99, SuarezGarcia14, Granero24, Kang14}. Such systems are often characterised by highly heterogeneous ~{\color{black} spatial or~{\color{black} temporal}} organisation in which relatively quiescent regions coexist with localised bursts of intense activity~\cite{Muzy06, Lakhal25}. Classical Gaussian models fail to reproduce this behaviour because they do not account for the strong scale-dependent variability of local fluctuation amplitudes~\cite{Jimenez2000, Friedrich11}. Multifractal and multiplicative stochastic models were introduced precisely to describe such intermittent organisation~\cite{Mandelbrot1974, Borlandetal2005}.

A central idea underlying many multifractal formulations is that observed fluctuations may be represented as rapidly varying stochastic activity modulated by a slower stochastic amplitude field. In this viewpoint, intermittency emerges not merely from fluctuations themselves, but from~{\color{black} spatial}ly varying modulation of fluctuation intensity across scales. The present work adopts this multiplicative modulation viewpoint as the conceptual foundation for localised wavelet-based unwrapping.

Importantly, the formulation developed below does not depend on a specific modulation construction. Instead, multiplicative cascades and multifractal random walks are treated as important stochastic realisations of a broader class of multiplicatively modulated multiscale processes. This distinction is essential for the localised interpretation developed in later sections.

\subsection{Multiplicative modulation fields}
\begin{figure}[htb]
    \centering
    \includegraphics[width=\linewidth]{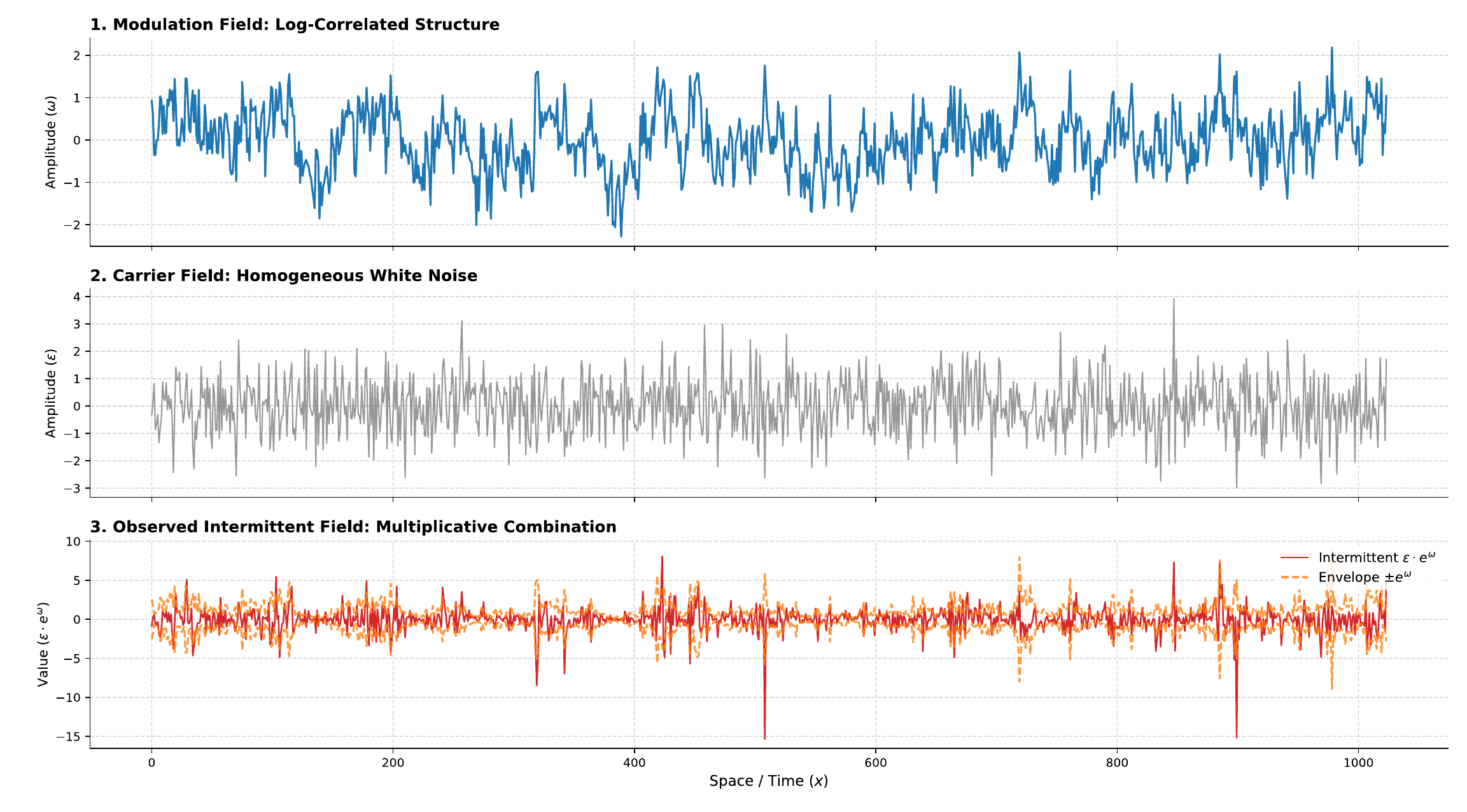}
%    \includesvg[width=\linewidth]{figure1.svg}
    \caption{From top to bottom: log-correlated carrier field $\omega(x)$, rapidly fluctuating modulation process $\epsilon(x)$ and observed intermittent field $X(x) = \epsilon(x)e^{\omega(x)}$ with indicated envelope $e^{\omega (x)}$.}\label{fig: envelopes}
\end{figure}

The central object considered in the present work is a multiplicatively modulated stochastic field of the form
\begin{equation}
X(x)=e^{\omega(x)}\epsilon(x),
\end{equation}
\noindent where:
\begin{align*}
    &\epsilon(x) \quad \text{denotes a rapidly fluctuating stochastic carrier process,} \\
    &\omega(x) \quad \text{denotes a stochastic modulation field controlling local fluctuation amplitudes.}
\end{align*}
\noindent
The observed process $X(x)$ combines two distinct forms of variability: local fluctuations carried by $\epsilon(x)$ and large-scale amplitude modulation governed by $\omega(x)$. The resulting intermittent field is therefore generated through local amplitude modulation of the carrier fluctuations by the multiplicative envelope $(e^{\omega(x)})$. This construction is illustrated schematically in Figure~\ref{fig: envelopes}.

The exponential representation is particularly natural because it guarantees positivity of the local amplitude modulation and transforms multiplicative interactions into additive relations under logarithmic transformation.
Conceptually, the modulation field controls the local strength of fluctuations. Regions where $\omega(x)$ is large correspond to enhanced local activity, while regions where $\omega(x)$ is small correspond to suppressed fluctuations. The resulting process therefore exhibits heterogeneous organisation across scales even when the carrier process itself remains comparatively simple.
This multiplicative viewpoint provides a natural framework for describing intermittency. The observed field may contain localised bursts, clustered singular structures, or strongly varying fluctuation intensity despite the underlying carrier process remaining statistically homogeneous. Such behaviour appears ubiquitously in multifractal systems.

The decomposition
\begin{equation}
X(x)=e^{\omega(x)}\epsilon(x)
\end{equation}
also immediately suggests the importance of logarithmic representations. Taking logarithms formally yields
\begin{equation}
\log |X(x)| = \omega(x)+\log |\epsilon(x)|.
\end{equation}
\noindent
This relation shows that logarithmic observables naturally separate modulation and carrier contributions additively. The present manuscript ultimately extends this principle into localised multiscale wavelet coordinates.

At this stage, however, the formulation remains completely general. No assumptions have yet been made regarding: (i)~the statistical structure of $\omega(x)$; (ii)~explicit modulation generation mechanisms; (iii)~exact scaling laws; or (iv)~particular stochastic constructions.
The framework therefore applies broadly to multiplicatively modulated multiscale processes independently of their detailed origin.

\subsection{MRW as a canonical stochastic realisation}

Among the most influential stochastic realisations of multiplicative modulation fields is the multifractal random walk (MRW) framework introduced by Bacry, Muzy, and collaborators~\cite{Bacryetal2000}. MRW models provide a statistically tractable formulation of intermittent multiscale processes while preserving essential multifractal characteristics~\cite{Bacryetal2000, Abryetal2009, Muzy06}.

In the MRW formulation, the modulation field $\omega(x)$ is typically modeled as a correlated Gaussian process possessing logarithmic covariance structure. The observed process retains the multiplicative form
\begin{equation}
X(x)=e^{\omega(x)}\epsilon(x),
\end{equation}
but the modulation field now satisfies specific scale-dependent statistical relations generating multifractal behaviour.
A central property of MRW models is the logarithmic covariance law
\begin{equation}
\mathrm{Cov}[\omega(x),\omega(x+r)] \sim -\lambda^2\log r,
\end{equation}
over an admissible scaling range. Here,
$\lambda^2$ denotes the intermittency parameter controlling the strength of multiscale modulation.
This logarithmic covariance structure plays a fundamental role in multifractal statistics. Slowly decaying logarithmic correlations induce scale-dependent fluctuation clustering and generate nontrivial scaling behaviour of moments and structure functions. In practice, the covariance of logarithmic observables provides one of the most robust signatures of multifractal modulation.

Importantly, the MRW framework does not require explicit discrete cascade trees. Although historically motivated by multiplicative cascade models, MRW admits a continuous stochastic formulation in which intermittency emerges through correlated modulation fields rather than through explicit branching constructions. This distinction becomes especially important in the localised wavelet interpretation developed later in the manuscript.

The work of Lakhal and collaborators further demonstrated that logarithmic covariance structure may be exploited operationally to estimate or “unwrap” modulation fields from observed intermittent processes. In such approaches, covariance estimation becomes a mechanism for probing hidden multiplicative organisation underlying observed fluctuations~\cite{Lakhal25}.
The present work builds directly on this perspective while reformulating the problem through localised wavelet representations.

\subsection{Logarithmic covariance structure and modulation extraction}
The multiplicative representation
\begin{equation}
X(x)=e^{\omega(x)}\epsilon(x)
\end{equation}
naturally leads to logarithmic observables. Formally,
\begin{equation}
\log |X(x)| = \omega(x)+\log |\epsilon(x)|.
\end{equation}
If the carrier fluctuations remain sufficiently decorrelated relative to the modulation field, the covariance of logarithmic observables becomes dominated by the covariance structure of
$\omega(x)$:
\begin{equation}
\mathrm{Cov} [ \log|X(x)|, \log|X(x+r)| ] \approx \mathrm{Cov} [ \omega(x), \omega(x+r) ].
\end{equation}
Under MRW assumptions this yields
\begin{equation}
\mathrm{Cov} [ \log|X(x)|, \log|X(x+r)| ] \sim -\lambda^2\log r.
\end{equation}

This relation provides the operational basis for covariance-based multifractal estimation. The logarithmic covariance structure acts as an observable statistical signature of multiplicative modulation.
Conceptually, covariance estimation may therefore be interpreted as an indirect mechanism for probing hidden modulation structure. Rather than observing the modulation field directly, one infers its statistical organisation through scale-dependent covariance properties of logarithmic observables.

The success of this strategy relies critically on approximate separability between:
the modulation field, and the carrier fluctuations. 
This separability is fundamentally statistical and global in classical MRW formulations. The covariance structure emerges after averaging over large ensembles or extended {\color{black}spatial} domains. Consequently, standard formulations provide limited direct access to localised modulation organisation. This limitation becomes particularly important in the presence of nonstationarity; heterogeneous intermittency; {\color{black}spatial}ly varying singular structures; localised defects; or finite-support observations. In such situations, global covariance estimation may obscure localised modulation behaviour. 

These considerations naturally motivate the introduction of localised multiscale representations capable of probing modulation structure simultaneously in scale 
and space.

\subsection{From global covariance statistics to localised multiscale probing}
Classical MRW and multifractal formulations are fundamentally global in nature. Statistical characterisation is typically performed through structure functions, partition functions, global covariance relations, or spectral averages.
Even wavelet-based multifractal methods such as WTMM ultimately use localised singularity information primarily to estimate global multifractal spectra through large-scale statistical aggregation.

The present work adopts a different perspective. Rather than using localisation solely as an intermediate step toward global statistics, we investigate whether multiplicative modulation itself may be probed locally through multiscale localised operators. This shift in viewpoint is central to the manuscript. The key question becomes: \textit{Can localised multiscale transforms create observation domains over which multiplicative modulation becomes approximately separable?}

Wavelet representations provide a natural framework for addressing this question because they simultaneously localise information in scale and position. Unlike Fourier representations, which remain globally distributed, wavelets define finite-support multiscale probing regions whose {\color{black}spatial} extent is explicitly controlled by the analyzing scale.
This localisation property suggests the possibility that sufficiently localised wavelets may effectively “freeze” the modulation field over their support, thereby enabling approximate local multiplicative separation.

The remainder of the manuscript develops this idea systematically. We show that wavelet localisation induces a localised modulation freezing approximation whose validity depends explicitly on support geometry, local regularity, and finite-scale multiscale mixing effects. This leads naturally to a localised operator reformulation of MRW unwrapping in wavelet coordinates.

\section{Wavelet Representation of Multiplicative Modulation}\label{sec3}
The multiplicative modulation framework introduced in the previous section provides a statistical description of intermittent multiscale organisation. However, the covariance-based formulations commonly employed in MRW analysis remain fundamentally global in character. Statistical estimation is typically performed through ensemble averages, long-range covariance functions, or global scaling relations. Such approaches characterise the modulation field indirectly through aggregated statistics rather than through explicitly localised observations. 
The present work adopts a different viewpoint based on localised multiscale probing. Instead of treating intermittency solely as a global statistical phenomenon, we investigate whether multiplicative modulation may be probed locally through multiscale operators possessing finite~{\color{black} spatial} support.

Wavelet transforms provide a natural framework for this purpose because they simultaneously localise information in both scale and position. Unlike Fourier representations, whose basis functions remain~{\color{black} spatial}ly delocalised, wavelets define localised observation domains whose~{\color{black} spatial} extent is explicitly controlled by the analyzing scale. This localisation property constitutes the essential operational mechanism underlying the present formulation.
Importantly, the role of the wavelet transform in the present context is not merely to provide a multiscale decomposition of the signal. Rather, wavelet localisation creates finite-support probing regions over which the modulation field may become approximately separable from the carrier fluctuations. The remainder of the manuscript develops the consequences of this observation systematically.

\subsection{Continuous wavelet transform and localised multiscale representation}
Let $\psi(x)$ denote an analyzing wavelet possessing sufficient regularity and vanishing moments. The continuous wavelet transform of a signal $X(x)$ is defined as
\begin{equation}
W_X(a,b) = \int X(x)\psi_{a,b}(x)\,dx,
\end{equation}
where
\begin{equation}
\psi_{a,b}(x) = \frac{1}{a^{d/2}} \psi\left( \frac{x-b}{a} \right), \label{eq:cwt wavelet definition}
\end{equation}
with $a>0$ denoting scale, $b$  denoting position, and $d$  representing dimensionality.

The wavelet coefficient $W_X(a,b)$ therefore measures the local behaviour of the signal around position $b$  at scale $a$. The transform simultaneously resolves:~{\color{black} spatial} localisation through $b$, scale localisation through $a$. 
This joint localisation distinguishes wavelet representations fundamentally from global spectral decompositions.
At small scales, the wavelet probes highly localised structures and fine-scale fluctuations. At larger scales, the support broadens and the transform becomes sensitive to progressively coarser organisation. The wavelet transform therefore defines a hierarchy of localised observation domains whose~{\color{black} spatial} extent evolves continuously with scale.

\begin{figure}[htb]
    \centering
    \includegraphics[width=\linewidth]{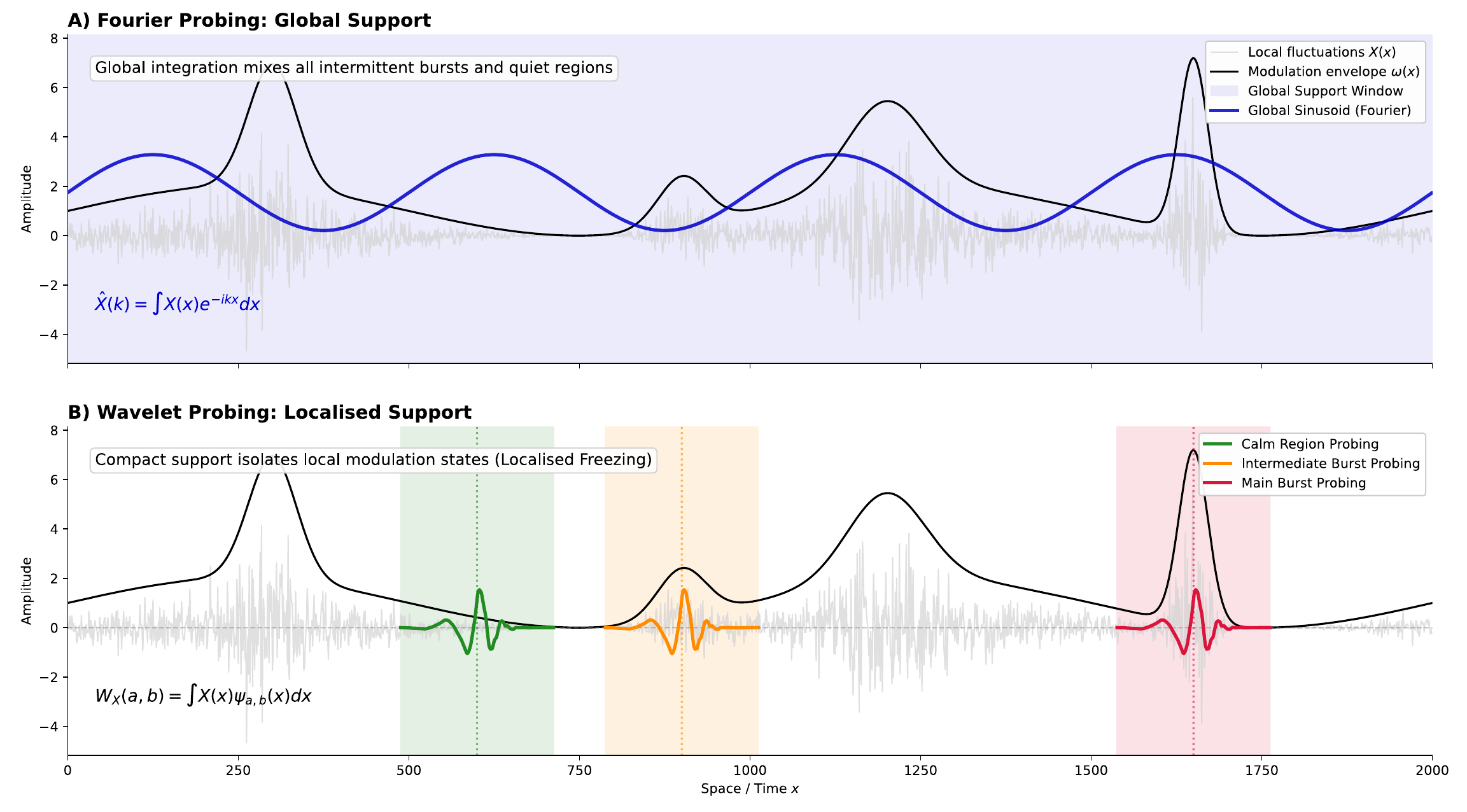}
%    \includesvg[width=\linewidth]{figure3.svg}
    \caption{Comparison of global Fourier and localised wavelet probing on an intermittent field $X(x)$ with an underlying modulation envelope $\omega(x)$. (A) The Fourier transform utilises sinusoidal basis functions with infinite, global support across the entire domain. While this yields perfect spectral resolution, the domain-wide integration inherently mixes quiet background regions with intense, intermittent bursts, completely blurring~{\color{black} temporal} or~{\color{black} spatial} locality. (B) In contrast, wavelet analysis employs basis functions $\psi_{a,b}(x)$ with strictly compact support. By translating the wavelet across varying~{\color{black} spatial} positions $b$, the integration is mathematically restricted to a confined local neighborhood (shaded regions). This prevents distant energy spikes from contaminating the local transform coefficient, cleanly isolating sharp, localised shocks and preserving the true~{\color{black} spatial} intermittency of the signal.}\label{fig: fourier_wavelet_comparison}
\end{figure}

Unlike Fourier basis functions, wavelets possess finite effective support and simultaneously localise information in scale and position. This localisation property allows wavelet coefficients to probe restricted neighbourhoods of the signal while remaining largely insensitive to distant structures. The geometric principles underlying wavelet localisation are illustrated in Figure~\ref{fig: wavelet_localisation} (Compare with Figure~\ref{fig: fourier_wavelet_comparison} for Fourier basis.).

Conceptually, the wavelet acts as a localised multiscale probe scanning the signal through finite-support observation windows. This interpretation is central to the present work.
In the context of multiplicative modulation fields,
\begin{equation}
X(x)=e^{\omega(x)}\epsilon(x),
\end{equation}
the wavelet coefficient becomes
\begin{equation}
W_X(a,b) = \int e^{\omega(x)} \epsilon(x) \psi_{a,b}(x)\,dx.
\end{equation}

At this stage no approximation has yet been introduced. However, the finite support of the wavelet immediately suggests the possibility that sufficiently localised wavelets may probe regions over which the modulation field varies only weakly. This observation forms the basis for the localised freezing approximation developed later in the manuscript.

Importantly, the present theoretical formulation is expressed entirely in continuous wavelet notation. This choice is deliberate. The continuous representation provides the natural analytical framework for describing localisation, support geometry, and local modulation behaviour independently of numerical discretisation.
The discrete implementation employed later should therefore be viewed as a computational realisation of the same localised probing principle rather than as the primary conceptual foundation.

\subsection{Wavelet support geometry and localised probing}
The essential property of the wavelet transform in the present context is not merely multiscale decomposition, but the creation of finite-support localised probing domains whose~{\color{black} spatial} extent is explicitly controlled by scale.
For compactly localised wavelets, the effective support of $\psi_{a,b}$ scales proportionally to $a$. The wavelet coefficient $W_X(a,b)$ therefore depends primarily on the behaviour of the signal within a neighbourhood centered around $b$ whose characteristic size is determined by the analyzing scale.

This support dependence is fundamental.
At fine scales, the probing region becomes highly localised and the transform primarily samples local signal organisation. At coarse scales, the support broadens and the coefficient integrates information over larger~{\color{black} spatial} domains. The wavelet transform thus defines a scale-dependent observation geometry.
This interpretation differs conceptually from classical spectral analysis. Fourier coefficients characterise global frequency content but do not define localised~{\color{black} spatial} neighbourhoods. Wavelet coefficients, by contrast, are intrinsically local objects whose support geometry is explicitly tied to scale.

In the present framework, this localisation property acquires direct operational significance. Because the modulation field
$\omega(x)$ is observed only through the finite support of the wavelet, the internal variability of the modulation process becomes support dependent.
More precisely, the wavelet coefficient probes the modulation field only over the effective support region
$x\in \mathrm{supp}(\psi_{a,b}).$
The finite support of the analysing wavelet implies that each coefficient is determined only by a restricted neighbourhood surrounding the analysis position. As illustrated schematically in Figure~\ref{fig: wavelet_localisation}D, the wavelet therefore acts as a local multiscale probe whose response is governed primarily by the structure contained within its support.

The quality of local multiplicative separation therefore depends on how strongly the modulation field varies within this neighbourhood. This immediately introduces a geometrical interpretation of scale: fine scales correspond to highly localised probing, whereas coarse scales correspond to increasingly averaged observations. The transition between these regimes becomes central for understanding both the validity and the limitations of localised MRW unwrapping.

The support geometry also naturally introduces finite-scale effects absent in purely global formulations. Since wavelet coefficients integrate over finite neighbourhoods, neighbouring structures inside the support may interact and generate residual multiscale coupling. Such effects become especially important at intermediate and coarse scales, where the support incorporates increasingly heterogeneous modulation structure.
As will be shown later, many of the covariance distortions observed numerically may be interpreted precisely as signatures of support-induced multiscale mixing.

Finally, the localised support interpretation provides a natural bridge toward local regularity theory. Since wavelet coefficients probe the signal over shrinking neighbourhoods as $a\to0$, their asymptotic behaviour becomes directly sensitive to local smoothness and singularity structure. This connection plays a fundamental role in the regularity-controlled freezing interpretation developed in Section~\ref{sec4}.

\subsection{Localisation versus global spectral representations}
The distinction between localised wavelet representations and global spectral methods is particularly important in the context of intermittent multiscale processes. Classical Fourier analysis decomposes signals into globally extended oscillatory components. The Fourier basis functions remain ~{\color{black}spatial}ly delocalised and therefore provide limited direct access to localised modulation structure. While spectral methods characterise global scaling properties efficiently, they do not naturally define finite observation neighbourhoods over which multiplicative modulation may become approximately separable. This distinction becomes especially critical for intermittent processes, heterogeneous singular structures, localised defects, nonstationary modulation, and ~{\color{black}spatial}ly varying regularity. In such situations, globally averaged spectral statistics may obscure local modulation behaviour.

Wavelet representations address this limitation through simultaneous localisation in scale and position. Each wavelet coefficient probes the signal through a finite-support observation domain whose geometry is explicitly controlled by the analyzing scale. The transform therefore provides localised access to multiscale organisation.

Importantly, the present work does not position wavelets merely as an alternative basis replacing Fourier methods. Rather, wavelet localisation fundamentally changes the observational structure of the problem itself. This essential shift can be framed as a transition from global statistical characterisation to localised multiscale probing:
\begin{equation}
    \text{Global Statistical Characterisation} \quad \longrightarrow \quad \text{Localised Multiscale Probing}.
\end{equation}
This distinction is central to the manuscript.
The wavelet transform does not simply reorganize information already available globally. Instead, localisation creates observation domains over which multiplicative modulation may become approximately frozen relative to the support of the analyzing wavelet. From this perspective, the wavelet transform acts not only as a representation tool but as a localised multiscale operator whose support geometry directly controls the validity of modulation extraction. This interpretation forms the conceptual basis for the localised modulation freezing approximation developed in the following section.

\subsection{Numerical realisation using compactly supported orthonormal wavelets}
The theoretical developments presented throughout this manuscript are formulated using continuous wavelet notation because the continuous representation provides the natural analytical language for describing localisation, support geometry, and local modulation behaviour. The numerical implementation employed in practice, however, relies on discrete orthonormal wavelet transforms constructed from compactly supported wavelet bases.

In the present work, dyadic orthonormal Daubechies wavelets are used for numerical realisation of the localised multiscale decomposition. These wavelets possess several properties particularly relevant for the present framework, namely compact ~{\color{black}spatial} support, multiscale localisation, orthonormality, and vanishing moments. Compact support is especially important because it preserves the localised probing interpretation underlying the freezing approximation. The wavelet coefficient remains primarily sensitive to signal organisation within a finite ~{\color{black}spatial} neighbourhood whose size scales dyadically with the analyzing scale.

The use of orthonormal wavelets also provides computational efficiency and numerical stability for large-scale stochastic simulations. Importantly, however, the discrete implementation should not be viewed as the conceptual foundation of the methodology. Rather, the discrete transform constitutes a computational approximation of the more general localised probing framework expressed continuously throughout the theoretical development.
This distinction between continuous analytical formulation and discrete numerical realisation is essential. The localised modulation freezing mechanism developed in the following section depends fundamentally on support geometry and localisation properties rather than on the specific discretisation itself.

Consequently, the theoretical interpretation remains applicable beyond the particular discrete implementation considered here and may naturally extend to alternative localised multiscale representations possessing suitable support localisation properties.

\begin{figure}[htb]
    \centering
    \includegraphics[width=\linewidth]{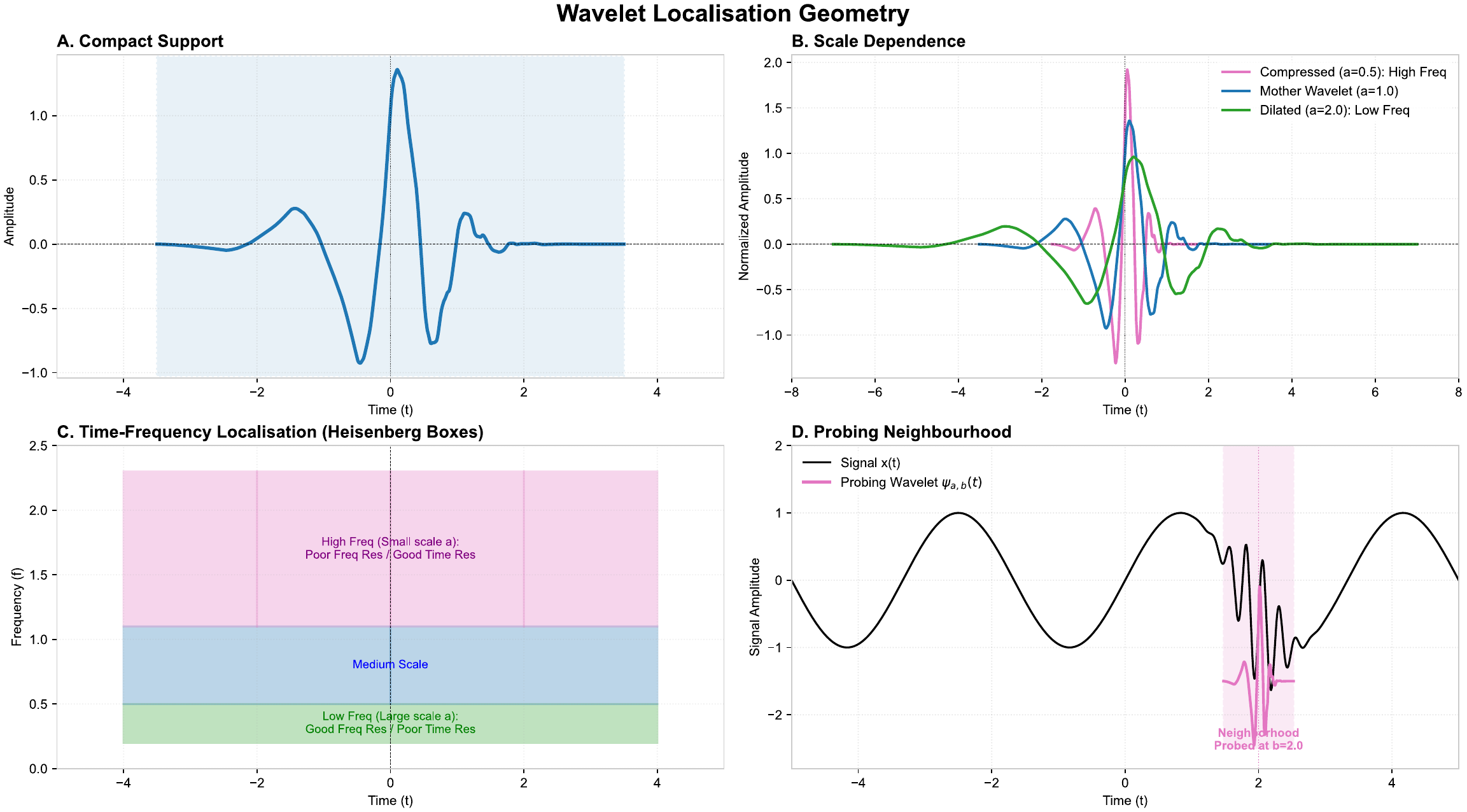}
%    \includesvg[width=\linewidth]{figure2.svg}
\caption{Wavelet localisation geometry and time-frequency analysis principles. (A) The Daubechies 4 ($\mathrm{db}4$) mother wavelet, $\psi(t)$, shifted to center its compact support around $t=0$. The shaded region highlights the finite~{\color{black} temporal} duration of the wavelet, a property that enables precise time-domain localisation. (B) Scale dependence of the wavelet family. Compressing the wavelet ($a=0.5$) isolates high-frequency components, while dilating it ($a=2.0$) captures low-frequency background trends. Amplitudes are scaled by a factor of $1/\sqrt{a}$ to conserve energy across all scales. (C) Time-frequency localisation limits represented via Heisenberg boxes. The varying aspect ratios illustrate the inherent resolution trade-off: large scales (green) provide excellent frequency resolution at the cost of~{\color{black} temporal} precision, whereas small scales (purple) offer the high~{\color{black} temporal} resolution necessary for pinpointing rapid transients. (D) Probing a specific neighborhood within a composite signal $x(t)$. By translating the wavelet to a specific time shift $b$ and applying a fine scale $a$, the probing wavelet $\psi_{a,b}(t)$ isolates a localised high-frequency anomaly, remaining completely unaffected by distant signal features outside its compact support.}\label{fig: wavelet_localisation}
\end{figure}

\section{Wavelet localisation and local modulation freezing}\label{sec4}
\subsection{Localised multiplicative wavelet representation}
The multifractal random walk (MRW) framework models intermittent fluctuations through a multiplicative modulation of an underlying stochastic carrier process. In its simplest form, the observed field may be written as
\begin{equation}X(x)=e^{\omega(x)}\epsilon(x),
\end{equation}
where
$\epsilon(x)$ denotes a rapidly fluctuating stochastic field and $\omega(x)$ is a correlated modulation process controlling the local amplitude structure. In classical MRW formulations, the statistical properties of $\omega(x)$  are primarily characterised through global covariance relations and scale-invariant stochastic laws. The present work adopts a complementary viewpoint based on localised multiscale probing of the modulation field through wavelet representations.
Let $\psi_{a,b}(x)$ denote a wavelet localised around position $b$ and scale $a$, with 
\begin{equation}
\psi_{a,b}(x)=\frac{1}{a^{d/2}}\psi\!\left(\frac{x-b}{a}\right),
\end{equation}
where $d$ denotes dimensionality. The corresponding wavelet coefficient of
\begin{equation}
W_X(a,b) = \int X(x)\psi_{a,b}(x)\,dx.
\end{equation}
Substituting the multiplicative representation of
$X(x)$ yields
\begin{equation}
W_X(a,b) = \int e^{\omega(x)} \epsilon(x) \psi_{a,b}(x)\,dx.
\end{equation}
At this stage, the role of wavelet localisation becomes fundamental. Because the wavelet is~{\color{black} spatial}ly localised, the integral probes the modulation field only within the effective support of $\psi_{a,b}$. If the internal variation of $\omega(x)$ over this support remains sufficiently weak, the modulation field may be approximated locally by its value at the wavelet centre:
\begin{equation}
\omega(x)\approx \omega(b), \qquad x\in \mathrm{supp}(\psi_{a,b}).
\end{equation}
Under this approximation, $e^{\omega(x)} \approx e^{\omega(b)}$, which can be factored outside the integral, yielding 
\begin{equation}
W_X(a,b) \approx e^{\omega(b)} \int \epsilon(x)\psi_{a,b}(x)\,dx.
\end{equation}
Therefore, 
\begin{equation}
W_X(a,b) \approx e^{\omega(b)}W_{\epsilon}(a,b).
\end{equation}
Taking logarithms of the modulus gives
\begin{equation}
\log \lvert W_X(a,b) \rvert \approx \omega(b)+\log |W_{\epsilon}(a,b)|.\label{eq:central}
\end{equation}
Equation~\eqref{eq:central} constitutes the central localised multiplicative approximation underlying the present wavelet formulation of MRW unwrapping. In contrast to global Fourier-domain formulations, the wavelet representation expresses the modulation process through localised multiscale probing of the modulation field. The approximation therefore does not rely on global stationarity, but instead on the local behaviour of the modulation field relative to the support of the analysing wavelet.

\begin{figure}[htb]
    \centering
    \includegraphics[width=\linewidth]{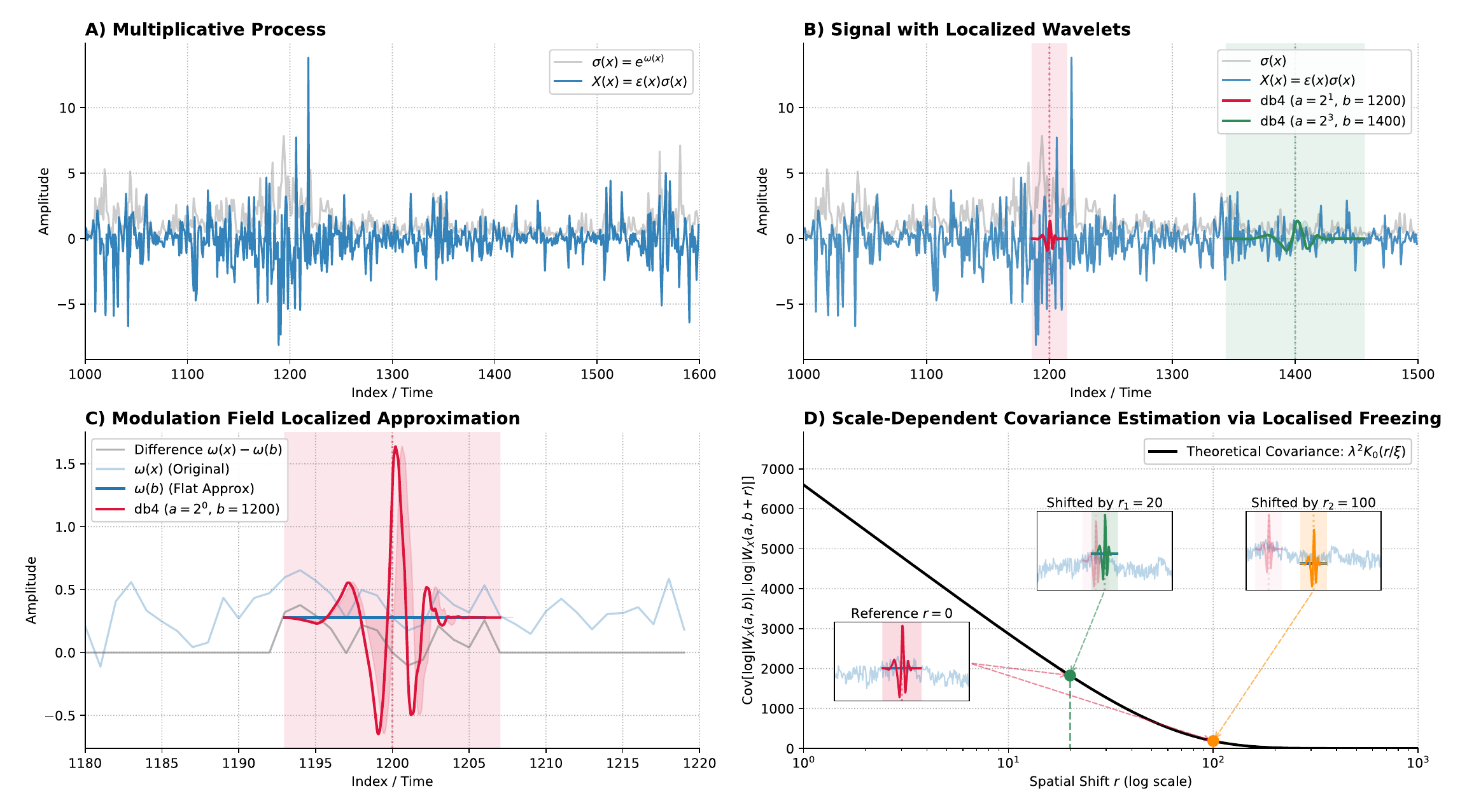}
%    \includesvg[width=\linewidth]{figure4.svg}
    \caption{Conceptual framework for extracting modulation statistics from an intermittent field. (A) An observed multiplicatively modulated signal, featuring fast local fluctuations $X(x)$ embedded within a heterogeneous, slowly varying stochastic envelope $e^{\omega(x)}$. (B) Localised wavelet probing isolates finite~{\color{black} spatial} neighborhoods, utilising a compact wavelet $\psi_{a,b}(x)$ at scale $a$ and translation $b$. (C) The principle of local modulation freezing. Within the wavelet's compact support, the modulation field is approximately constant, $\omega(x) \approx \omega(b)$, allowing the logarithmic wavelet amplitudes to cleanly separate as $\log\lvert W_X(a,b) \rvert \approx \omega(b) + \log|W_\epsilon(a,b)|$. (D) Idealised localised covariance extraction. Correlating these locally frozen wavelet amplitudes across a~{\color{black} spatial} lag $r$ recovers the underlying theoretical modulation covariance, yielding a characteristic $-\lambda^2\log r$ scaling prior to the onset of finite-scale deviations.}\label{fig: conceptual_framework}
\end{figure}

Conceptually, the wavelet acts here as a localised multiscale probe whose support defines an observation neighbourhood over which the modulation field may become effectively frozen. At sufficiently fine scales, the modulation process varies slowly relative to the wavelet support, allowing the multiplicative modulation to behave locally as an approximately constant amplitude envelope acting on the stochastic carrier field. In this regime, wavelet localisation induces what may be termed local modulation freezing. 
{\color{black} The complete operational framework is summarised schematically in Figure~\ref{fig: conceptual_framework}. Beginning from an observed multiplicatively modulated field, wavelet localisation generates finite-support probing domains within which the modulation field may be treated as approximately frozen, thereby enabling local covariance-based recovery of modulation statistics.}
Importantly, the approximation is asymptotic rather than exact. Its validity depends explicitly on the internal variability of $\omega(x)$ within the wavelet support. As the analysing scale increases, or as the modulation field becomes increasingly irregular, the local freezing approximation progressively deteriorates and additional multiscale coupling terms emerge. These residual interactions are responsible for the finite-scale covariance distortions observed numerically in later sections.

\subsection{Local modulation freezing approximation}
\begin{figure}[htb]
    \centering
    \includegraphics[width=0.9\linewidth]{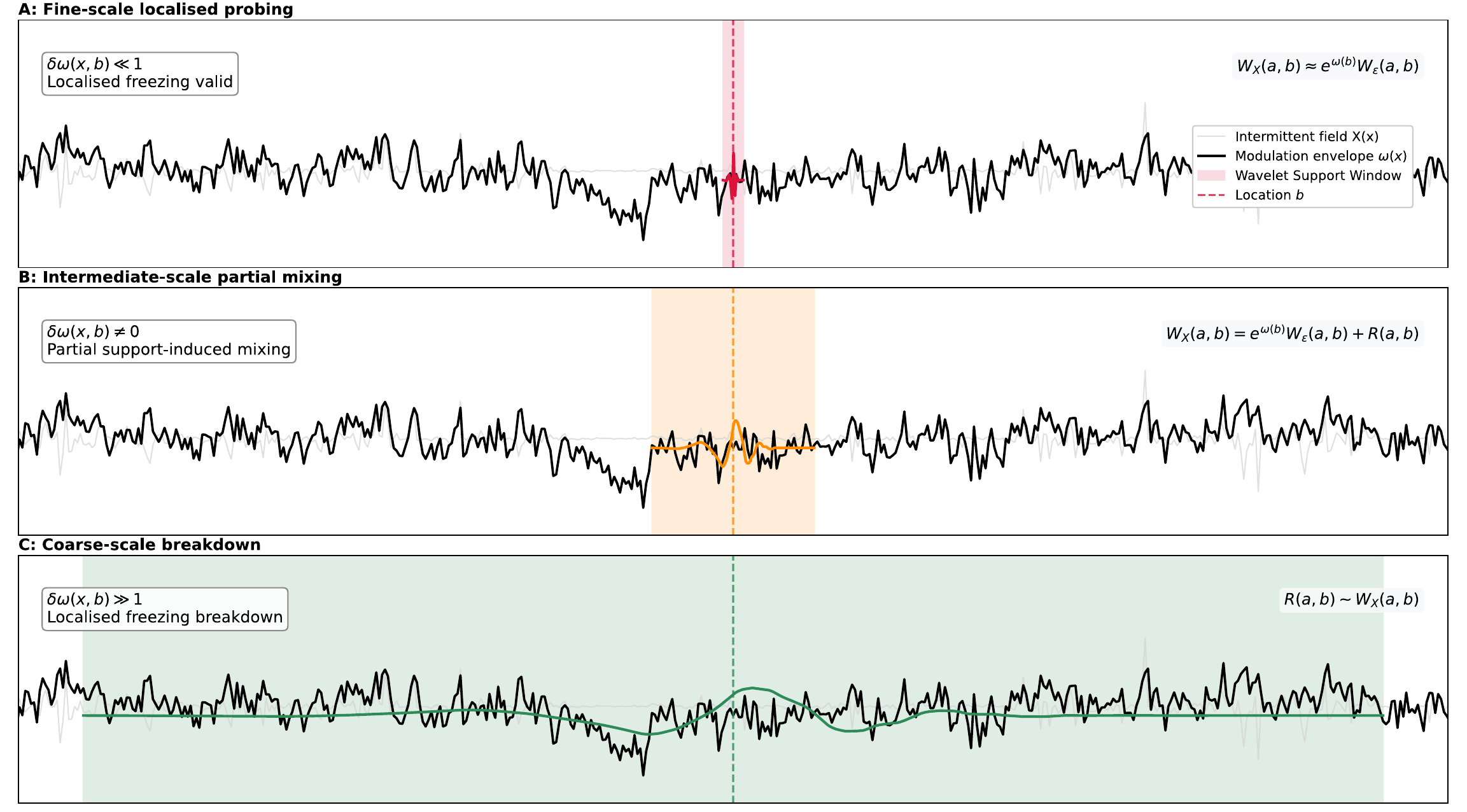}
%    \includesvg[width=0.9\linewidth]{figure5.svg}
    \caption{A geometric illustration of why the validity of local modulation freezing is fundamentally dictated by the scale of observation. All panels display the identical underlying intermittent signal and its stochastic modulation envelope $\omega(x)$, varying only the wavelet support size to isolate geometric effects. (A) Fine-scale localised probing. The compact wavelet support operates within a nearly homogeneous modulation neighborhood ($\delta\omega(x,b) \ll 1$), perfectly preserving the idealised multiplicative separation $W_X(a,b) \approx e^{\omega(b)}W_\epsilon(a,b)$. (B) Intermediate-scale partial mixing. As the support expands, it begins to span visibly different modulation states ($\delta\omega(x,b) \neq 0$), integrating over multiple local regimes and giving rise to a residual mixing term $R(a,b)$. (C) Coarse-scale breakdown. The wavelet support overextends across massive envelope heterogeneity ($\delta\omega(x,b) \gg 1$), resulting in a complete failure of the freezing approximation as support-induced mixing dominates ($R(a,b) \sim W_X(a,b)$). This scale-dependent integration conceptually predicts the inevitable emergence of finite-scale covariance distortions.}\label{fig: modulation_freezing}
\end{figure}

The localised multiplicative representation derived in the previous subsection suggests that wavelet localisation induces an effective decoupling between the modulation field and the stochastic carrier process. This approximation, however, is not exact and requires a more precise characterisation of the conditions under which it remains valid. In the present framework, this motivates the introduction of the concept of local modulation freezing.
Consider again the multiplicative representation
\begin{equation}
X(x)=e^{\omega(x)}\epsilon(x),
\end{equation}
together with the wavelet coefficient
\begin{equation}
W_X(a,b) = \int e^{\omega(x)} \epsilon(x)\psi_{a,b}(x)\,dx.
\end{equation}
Let the modulation field be decomposed locally as
\begin{equation}
\omega(x) = \omega(b) + \delta\omega(x,b),\label{eq: approximation}
\end{equation}
where
\begin{equation}
\delta\omega(x,b) = \omega(x)-\omega(b)
\end{equation}
denotes the internal variation of the modulation field relative to the wavelet centre $b$. Substituting into the wavelet representation yields
\begin{equation}
W_X(a,b) = e^{\omega(b)} \int e^{\delta\omega(x,b)} \epsilon(x)\psi_{a,b}(x)\,dx.
\end{equation}
The localised multiplicative approximation introduced corresponds to neglecting the internal fluctuation term $\delta\omega(x,b)$ over the effective support of the wavelet. {\color{black} {This is illustrated in Figure~\ref{fig: modulation_freezing}C --- sufficiently localised wavelet supports permit the approximation $\omega(x)\approx\omega(b)$, within the analysis neighbourhood, yielding the factorised representation of Eqn.~\ref{eq: approximation}.}}
This motivates the following operational definition.

\begin{definition}[\bf local modulation freezing]
Let
$\psi_{a,b}$
denote a wavelet localised at scale $a$  and position $b$. The multiplicative modulation field is said to exhibit local modulation freezing at $(a,b)$ if
\begin{equation}
|\delta\omega(x,b)| \ll 1
\end{equation}
throughout the effective support of $\psi_{a,b}$, yielding the approximation
\begin{equation}
W_X(a,b) \approx e^{\omega(b)}W_{\epsilon}(a,b).
\end{equation}
\end{definition}

The approximation becomes asymptotically valid as the internal variability of the modulation field within the wavelet support tends to zero.
This definition emphasizes that freezing is fundamentally a relative and scale-dependent concept. The modulation field is not assumed to be globally constant, smooth, or stationary. Rather, the wavelet support defines a localised observation domain over which the modulation process may become effectively frozen relative to the analysing scale.
This distinction is essential. In classical multifractal cascade models, the modulation field may remain highly irregular globally while still appearing locally coherent when probed by sufficiently localised wavelets. The freezing approximation therefore reflects not the absence of intermittency, but rather the finite~{\color{black} spatial} extent of the multiscale probe itself.

The dependence on wavelet support is central to this interpretation. Since the support of $\psi_{a,b}$ scales proportionally to $a$, decreasing the analysing scale reduces the~{\color{black} spatial} region over which the modulation field must remain approximately constant. Consequently, the freezing approximation naturally improves toward finer scales:
\begin{equation}
a\rightarrow 0.
\end{equation}

In this asymptotic regime, the wavelet increasingly probes the modulation field locally rather than globally, and the multiplicative modulation becomes progressively separable from the carrier fluctuations. 
The validity of the local freezing approximation is therefore not universal, but depends explicitly on the interaction between wavelet support geometry and internal modulation variability. As the support expands, progressively larger modulation fluctuations become incorporated into a single coefficient, eventually violating the assumptions underlying local factorisation. This geometric mechanism is illustrated schematically in Figure~\ref{fig: modulation_freezing}.
Conceptually, the wavelet acts here as a localised microscope observing the modulation field over a finite~{\color{black} spatial} neighbourhood. At coarse scales, the microscope integrates over broad regions containing substantial internal modulation variability. At fine scales, the observation window contracts and the modulation field becomes increasingly homogeneous within the wavelet support. Local modulation freezing therefore emerges as a direct consequence of multiscale localisation.

This interpretation provides an intuitive explanation for the scale dependence observed in numerical covariance estimation. Fine-scale wavelet coefficients probe regions where the modulation field remains approximately frozen, leading to covariance structures consistent with theoretical predictions. As the scale increases, internal fluctuations within the wavelet support become progressively non-negligible, generating residual multiscale coupling and distortions of the idealised covariance behaviour.
Importantly, local modulation freezing does not imply exact scale decoupling. Even when the approximation is valid asymptotically, finite-support effects generate residual interactions between modulation variability and the stochastic carrier process. These interactions become particularly important at intermediate and coarse scales, where the internal structure of the modulation field can no longer be neglected. The resulting multiscale mixing effects will be analysed explicitly in the following subsection.

The freezing viewpoint also clarifies the role of wavelets within the present reformulation of Lakhal-style MRW unwrapping. The purpose of the wavelet transform is not merely to decompose the signal into scales, but to create localised observation domains over which the multiplicative modulation may become approximately frozen. Wavelet localisation therefore provides the operational mechanism enabling localised extraction of the modulation field from multiscale coefficients.
This perspective differs fundamentally from classical Fourier-domain formulations. Fourier representations provide global spectral decomposition but do not naturally define localised~{\color{black} spatial} neighbourhoods over which the modulation field may be approximated as locally constant. In contrast, wavelet localisation introduces a direct connection between analysing scale,~{\color{black} spatial} support, and local modulation structure. The validity of the unwrapping procedure thus becomes explicitly controlled by the relation between wavelet support and the internal variability of the modulation field.

Finally, the local modulation freezing approximation naturally connects the present framework with wavelet-based regularity theory. The validity of freezing depends on the local variability of $\omega(x)$ within the wavelet support, suggesting that local regularity properties of the modulation field govern the asymptotic accuracy of the approximation. This connection provides the basis for the regularity-controlled interpretation developed in subsequent sections. 

\subsection{Residual multiscale mixing}
The local modulation freezing approximation provides a natural mechanism through which multiplicative modulation becomes approximately separable in wavelet space. However, the approximation is intrinsically finite-scale and therefore necessarily imperfect. The residual deviations observed numerically in covariance estimation arise precisely from the breakdown of exact freezing within the wavelet support. In the present framework, these effects are interpreted as manifestations of residual multiscale mixing.

Starting from the local decomposition
\begin{equation}
\omega(x) = \omega(b)+\delta\omega(x,b),
\end{equation}
the wavelet coefficient may be written as
\begin{equation}
W_X(a,b) = e^{\omega(b)} \int e^{\delta\omega(x,b)} \epsilon(x)\psi_{a,b}(x)\,dx.
\end{equation}
Under ideal freezing conditions,
\begin{equation}
\delta\omega(x,b)\approx 0,
\end{equation}
and the multiplicative approximation reduces to
\begin{equation}
W_X(a,b) \approx e^{\omega(b)}W_\epsilon(a,b).
\end{equation}
At finite scales, however, the internal variability term $\delta\omega(x,b)$ remains non-negligible. To analyse its effect, consider the local exponential expansion
\begin{equation}
e^{\delta\omega(x,b)} = 1 + \delta\omega(x,b) + \frac{1}{2}\delta\omega(x,b)^2 + \cdots
\end{equation}
Substituting into the wavelet representation yields
\begin{equation}
W_X(a,b) = e^{\omega(b)} \left[ W_\epsilon(a,b) + R_1(a,b) + R_2(a,b) + \cdots \right],
\end{equation}
where
\begin{equation}
R_1(a,b) = \int \delta\omega(x,b)\, \epsilon(x)\, \psi_{a,b}(x)\,dx
\end{equation}
and
\begin{equation}
R_2(a,b) = \frac12 \int \delta\omega(x,b)^2 \epsilon(x) \psi_{a,b}(x)\,dx.
\end{equation}
The terms $R_n(a,b)$ represent residual multiscale coupling contributions generated by internal variability of the modulation field within the wavelet support. These terms vanish only in the idealised asymptotic freezing limit. At finite scales, they introduce additional interactions between the modulation field and the stochastic carrier process.

The essential point is that the wavelet coefficient no longer depends solely on the local value $\omega(b)$, but also on the~{\color{black} spatial} organisation of fluctuations inside the support of the analysing wavelet. The coefficient therefore acquires a nonlocal internal structure reflecting unresolved modulation variability.
This mechanism naturally explains the covariance distortions observed numerically. Under ideal freezing conditions,
\begin{equation}
\log\lvert W_X(a,b) \rvert \approx \omega(b) + \log|W_\epsilon(a,b)|,
\end{equation}
and the covariance structure is dominated by the statistics of the modulation field:
\begin{equation}
\mathrm{Cov} \left[ \log\lvert W_X(a,b) \rvert, \log|W_X(a,b+r)| \right] \sim \mathrm{Cov}[\omega(b),\omega(b+r)].
\end{equation}
For MRW-type modulation fields this yields the expected logarithmic covariance behaviour.

Residual multiscale mixing modifies this structure. The observed covariance now contains additional finite-scale contributions originating from the coupling terms $R_n(a,b)$. Formally, one may write
\begin{equation}
\log\lvert W_X(a,b) \rvert = \omega(b) + \log|W_\epsilon(a,b)| + M(a,b),
\end{equation}
where $M(a,b)$  denotes the effective multiscale mixing contribution generated by internal modulation variability.

The covariance therefore becomes
\begin{equation}
\mathrm{Cov} \left[ \log\lvert W_X(a,b) \rvert, \log|W_X(a,b+r)| \right] = C_\omega(r) + C_{\mathrm{mix}}(a,r) + C_{\epsilon}(a,r),
\end{equation}
\noindent where $C_\omega(r)$ denotes the ideal modulation covariance contribution, $C_{\mathrm{mix}}(a,r)$ represents residual multiscale mixing, and $C_{\epsilon}(a,r)$ contains carrier-dependent finite-scale terms.
The mixing contribution
$C_{\mathrm{mix}}(a,r)$
constitutes the central finite-scale correction mechanism in the present framework. Importantly, it depends explicitly on both scale and lag:
\begin{equation}
C_{\mathrm{mix}}=C_{\mathrm{mix}}(a,r).
\end{equation}
This dependence reflects the fact that residual mixing is generated by the interaction between wavelet support geometry and the internal~{\color{black} spatial} organisation of the modulation field.
Conceptually, multiscale mixing emerges because the wavelet coefficient integrates over a finite~{\color{black} spatial} neighbourhood rather than probing an infinitesimal point. At sufficiently fine scales, the support becomes small enough for the modulation field to appear approximately frozen, and the mixing contribution remains weak. As the analysing scale increases, the support progressively incorporates larger internal modulation variability, causing the residual coupling terms to grow.

This interpretation provides a direct explanation for the characteristic finite-scale structures observed numerically in covariance estimation. In particular, the secondary covariance bumps appearing at intermediate lags may be understood as signatures of effective support-induced modulation mixing. These structures do not necessarily indicate numerical instability or algorithmic failure. Rather, they reflect the progressive deterioration of local freezing as the analysing wavelet integrates over increasingly heterogeneous modulation regions.

\begin{figure}[htb]
    \centering
    \includegraphics[width=\linewidth]{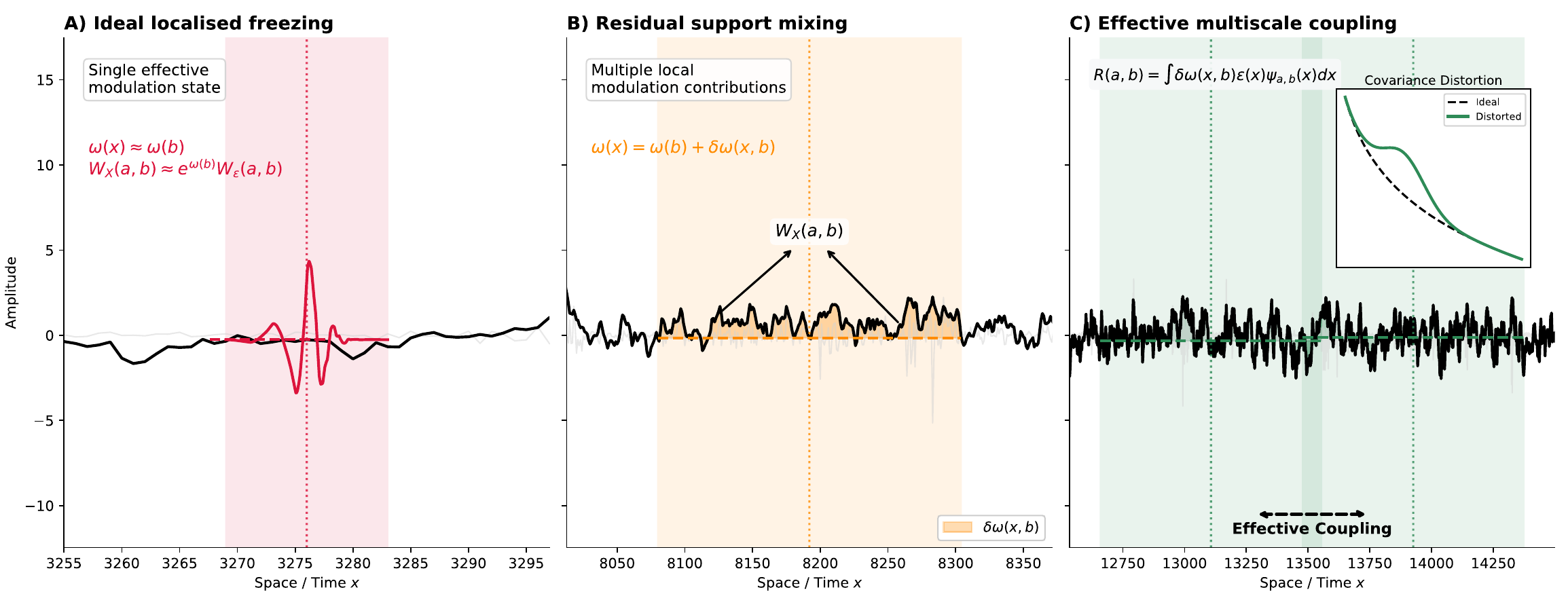}
%    \includesvg[width=\linewidth]{figure6.svg}
    \caption{Mechanistic origin of finite-scale covariance distortions arising from internal modulation variability within the wavelet support. (A) Ideal localised freezing. A compact wavelet captures a single effective modulation state ($\omega(x) \approx \omega(b)$), yielding pure multiplicative separation without residual coupling. (B) Residual support mixing. When the support spans heterogeneous regions, the local envelope decomposes into $\omega(x) = \omega(b) + \delta\omega(x,b)$. The resulting wavelet coefficient integrates these multiple varying contributions, generating support-induced mixing. (C) Effective multiscale coupling. Residual internal variability introduces an effective interaction between nearby supports, formalised by the mixing operator $R(a,b) = \int \delta\omega(x,b) \epsilon(x)\psi_{a,b}(x) dx$. This non-local coupling mechanism fundamentally distorts the idealised logarithmic covariance structure, physically predicting the theoretically anticipated finite-scale deviations.}\label{fig: covariance_origin}
\end{figure}

\subsection{Local regularity interpretation}
The local modulation freezing approximation introduced above suggests that the validity of localised multiplicative factorisation is fundamentally controlled by the internal variability of the modulation field within the wavelet support. This naturally connects the present framework with wavelet-based local regularity theory, where wavelet coefficients characterise the asymptotic behaviour of singular or irregular structures across scales~\cite{MallatHwang1992, Muzyetal1991}. Wavelet regularity analysis is classically based on the observation that local Hölder regularity determines the asymptotic scaling of wavelet coefficients~\cite{daubechies92, Mallat1999}.

Let $f(x)$ belong locally to the Hölder space $C^h(x_0)$ around a point $x_0$. Then, under standard assumptions on the analysing wavelet,
\begin{equation}
    |W_f(a,b)| \sim a^h, \qquad b\to x_0, \qquad a\to0.
\end{equation}
This fundamental result, developed extensively in wavelet regularity theory and particularly in the work of \text{\textit{Stéphane Jaffard}}~\cite{Jaffard1997a, Jaffard1997b}, establishes wavelet coefficients as localised probes of singularity structure and local smoothness~\cite{MallatHwang1992, Mallat1999}.

In the present context, the relevant object is not directly the carrier process $\epsilon(x)$, but rather the modulation field $\omega(x)$. The freezing approximation depends on the local variability
\begin{equation}
\delta\omega(x,b) = \omega(x)-\omega(b)
\end{equation}
within the effective support of the wavelet. If $\omega(x)$ possesses local Hölder regularity exponent $h_\omega(b)$, then
\begin{equation}
|\delta\omega(x,b)| \lesssim |x-b|^{h_\omega(b)}
\end{equation}
locally around $b$. Since the support diameter of the wavelet scales proportionally to $a$, one obtains
\begin{equation}
|\delta\omega(x,b)| \lesssim a^{h_\omega(b)}.
\end{equation}

Consequently, the internal variability controlling the freezing approximation decreases asymptotically according to the local regularity of the modulation field itself. Once the wavelet support spans multiple modulation states, the modulation field can no longer be treated as constant within the analysis neighbourhood. The resulting partial support averaging generates the residual contribution R(a,b), illustrated schematically in Figure~\ref{fig: modulation_freezing}B and in Figure~\ref{fig: covariance_origin}C.
This directly yields a regularity-controlled estimate for the residual multiscale mixing introduced in the previous subsection. Since the leading correction term is
\begin{equation}
R_1(a,b) = \int \delta\omega(x,b)\, \epsilon(x)\, \psi_{a,b}(x)\,dx,
\end{equation}
its amplitude is asymptotically controlled by the local variation of $\omega(x)$:
\begin{equation}
R_1(a,b) \sim a^{h_\omega(b)}.
\end{equation}
More generally, the effective multiscale mixing contribution satisfies
\begin{equation}
M(a,b) \sim a^{h_\omega(b)}
\end{equation}
in the asymptotic fine-scale limit.

This relation provides a direct connection between local regularity and the validity of localised modulation freezing. Regions where the modulation field is smoother admit stronger local freezing and weaker multiscale mixing. Conversely, highly irregular regions generate stronger residual coupling effects and earlier breakdown of the freezing approximation. In the large-scale regime, internal modulation variability may become comparable to or larger than the carrier fluctuations themselves. In this limit the residual term is no longer perturbative and the freezing approximation ceases to provide an accurate local representation (Figure~\ref{fig: modulation_freezing}C).
Conceptually, the wavelet does not merely observe the modulation field; it interrogates its local regularity structure. The quality of multiplicative separation therefore depends not only on scale, but also on the local geometric organisation of the modulation field itself.
This interpretation provides a natural explanation for the~{\color{black} spatial} heterogeneity often observed numerically in localised covariance estimation and reconstruction quality. Different regions of the modulation field may exhibit different effective freezing behaviour depending on the local regularity properties of $\omega(x)$. The validity of localised unwrapping is therefore intrinsically nonuniform in across the analysis domain.

Importantly, the present viewpoint differs from classical multifractal regularity analysis in a fundamental way. In standard wavelet multifractal formalisms, local Hölder exponents are primarily used to construct global multifractal spectra through partition-function averaging procedures~\cite{Muzyetal1991, Muzyetal1993, Lashermesetal2004}. Local regularity acts there as an intermediate quantity whose final role is statistical characterisation of singularity populations~\cite{FrischParisi1985, Jaffard1997a, Wendtetal2007}.
Here, by contrast, local regularity directly controls the operational validity of localised modulation probing itself. The Hölder structure of the modulation field determines the asymptotic accuracy of the freezing approximation and therefore the degree to which multiplicative modulation may be locally separated from the carrier fluctuations.

The present framework thus introduces a different operational role for wavelet regularity theory. Rather than using wavelets solely to characterise singularity distributions, wavelet localisation becomes the mechanism through which local modulation may be approximately unfolded.
This distinction is particularly important in relation to earlier wavelet-transform modulus maxima (WTMM) approaches~\cite{Muzyetal1991, Muzyetal1993}. In WTMM theory, singularities are tracked geometrically through maxima lines propagating across scales, ultimately yielding global multifractal spectra~\cite{MallatHwang1992, Muzyetal1993}. In the present framework, the focus shifts from singularity census toward localised operator behaviour. The central question is no longer only \textit{“What singularities are present?”} but rather \textit{“Over what scales does local multiplicative freezing remain valid?”} This shift from global statistical characterisation toward localised modulation interrogation constitutes one of the central conceptual differences of the present approach.

The regularity interpretation also clarifies the asymptotic nature of wavelet-based unwrapping. Since the freezing approximation improves according to $a^{h_\omega(b)}$, the method is intrinsically controlled by the fine-scale limit. Exact reconstruction is not expected at finite scales, particularly in regions where the modulation field becomes strongly irregular. Instead, the methodology should be interpreted as an asymptotically localised approximation whose operational validity depends on the interplay between scale, support geometry, and local regularity.

Finally, this regularity-controlled perspective naturally explains the scale-dependent transition observed numerically between freezing-dominated and mixing-dominated regimes. At sufficiently fine scales, local Hölder regularity suppresses internal modulation variability and localised factorisation becomes increasingly accurate. At coarser scales, unresolved modulation variability accumulates within the wavelet support, generating progressively stronger multiscale coupling and covariance distortion. The observed breakdown of ideal logarithmic covariance behaviour may therefore be interpreted as a direct manifestation of finite-scale regularity limitations in localised modulation probing.

\subsection{Consequences for localised MRW unwrapping}
The localised modulation freezing framework developed above has direct consequences for Lakhal-style MRW unwrapping and, more generally, for localised multiscale reconstruction of multiplicative modulation. In the present interpretation, the wavelet representation is not merely a numerical reformulation of an existing procedure, but a localised operator framework whose validity is explicitly controlled by wavelet support geometry and local modulation regularity. Classical Lakhal-style unwrapping is fundamentally based on the logarithmic covariance structure of multiplicative modulation models. In its original formulation, the modulation field is inferred statistically through scale-dependent covariance relations derived from globally defined representations. The wavelet formulation developed here preserves this conceptual foundation while fundamentally modifying the observational geometry of the problem. Specifically, the wavelet transform introduces localised multiscale probing domains over which the modulation field may become approximately frozen. Under the freezing approximation,
\begin{equation}
W_X(a,b) \approx e^{\omega(b)}W_\epsilon(a,b),
\end{equation}
yielding
\begin{equation}
\log\lvert W_X(a,b) \rvert \approx \omega(b) + \log|W_\epsilon(a,b)|.
\end{equation}

This relation implies that the covariance structure of logarithmic wavelet amplitudes contains direct localised information regarding the modulation field itself. The wavelet-domain formulation therefore transforms MRW unwrapping into a localised multiscale estimation problem.

Importantly, the present framework reveals that the validity of this estimation procedure depends explicitly on scale. The freezing approximation improves asymptotically toward fine scales, where the wavelet support becomes sufficiently localised relative to the internal variability of the modulation field. Conversely, at coarse scales the wavelet integrates over increasingly heterogeneous regions, producing residual multiscale mixing and progressive degradation of covariance separability. 

A central practical consequence emerges immediately from this behavior: scale selection is intrinsic to the validity of localised unwrapping. In classical global covariance formulations, scaling ranges are typically selected statistically through approximate linearity of covariance curves. In the present framework, however, the scaling range acquires a direct operational interpretation. The admissible scales are precisely those for which local modulation freezing remains sufficiently valid.

The scale-selection problem therefore becomes geometrically and physically interpretable rather than purely empirical. The upper scale limit is determined by the onset of significant internal modulation variability within the wavelet support, while the lower limit may be constrained by discretisation effects, finite sampling, carrier noise, or wavelet-dependent support artifacts.
This interpretation also clarifies the numerical covariance distortions observed in the preceding sections. Residual covariance bumps, crossover regions, and loss of apparent logarithmic scaling do not necessarily indicate algorithmic failure. Instead, they represent measurable signatures of finite-scale breakdown of local freezing.

From this perspective, covariance deviations become informative rather than purely undesirable. They indicate precisely where the localised approximation begins to deteriorate and where multiscale mixing becomes dominant. The transition between freezing-dominated and mixing-dominated regimes therefore defines the operational validity range of wavelet-based unwrapping.
The present framework furthermore suggests the possibility of adaptive localised fitting strategies. Since freezing validity depends on local regularity and support-dependent mixing, different regions of the signal or image may admit different admissible scaling ranges. Rather than imposing globally fixed fitting intervals, one may instead estimate localised validity domains determined by the effective strength of residual multiscale mixing.

Conceptually, this leads to an important reinterpretation of MRW unwrapping itself. The goal is no longer exact inversion of a globally defined multiplicative modulation field, but rather asymptotically localised extraction of modulation structure over scales where modulation freezing remains approximately valid.
This distinction is fundamental. Exact global inversion of multiplicative modulation fields is generally ill-posed due to the intrinsically multiscale and stochastic nature of the modulation process. The present framework instead introduces a localised operator viewpoint in which reconstruction quality is controlled explicitly by support geometry, local regularity, and scale-dependent mixing.

The wavelet formulation also naturally explains why compactly supported orthonormal wavelets provide a viable computational realisation of the methodology despite the continuous formulation adopted theoretically. Local freezing depends primarily on support localisation rather than on exact transform continuity. Dyadic orthonormal wavelets therefore preserve the essential operational mechanism provided that the support remains sufficiently localised relative to the modulation variability.
This perspective resolves an important conceptual tension between continuous analytical formulations and discrete numerical implementations. The continuous wavelet representation provides the natural theoretical language for expressing local modulation freezing and multiscale localisation, while the discrete wavelet transform supplies a computational approximation of the same localised probing mechanism.

The present interpretation also suggests that residual finite-scale effects should not merely be suppressed numerically, but characterised structurally. Since covariance distortions arise from effective multiscale mixing kernels induced by localised probing, these deviations may themselves contain information regarding the interaction between modulation field organisation and wavelet support geometry.

This observation opens several possible future directions, including: (i)~adaptive support-dependent unwrapping; (ii)~regularity-controlled scale selection; (iii)~multiscale validity diagnostics; (iv)~localised modulation field segmentation; and (v)~support-corrected covariance estimation. Such extensions lie beyond the scope of the present work, but emerge naturally from the localised operator interpretation developed here.

Finally, the present framework clarifies the broader conceptual role of wavelets in multiplicative modulation analysis. Wavelets are not introduced merely as an alternative multiscale representation replacing Fourier methods. Rather, wavelet localisation fundamentally changes the observational structure of the problem by creating localised probing domains within which multiplicative modulation may become asymptotically frozen.
In this sense, wavelet localisation acts as the operational mechanism enabling localised multiscale interrogation of multiplicative modulation. 

The resulting reformulation of Lakhal-style MRW unwrapping therefore represents not simply a change of basis, but a transition from global statistical characterisation toward localised modulation probing.
Within this framework, wavelet localisation no longer acts solely as a passive decomposition principle. Instead, localised freezing transforms localisation into an active multiscale probing mechanism whose validity depends on support geometry, finite-scale overlap, and local modulation variability. {\color{black} The theoretical development presented above suggests that finite-scale covariance distortions should emerge naturally from residual support mixing rather than from a breakdown of the underlying multiplicative model itself. The proposed mechanism is summarised in Figure~\ref{fig: covariance_origin}. and the present section develops this framework systematically.}

\section{Localised Wavelet Formulation of MRW Unwrapping}\label{sec5}
The theoretical developments presented in the preceding sections suggest that wavelet localisation induces approximate local multiplicative separation over finite-support probing domains. The purpose of the present section is to operationalise this framework into a localised formulation of MRW unwrapping and covariance estimation. This operational shift means that while classical MRW unwrapping operates primarily through global covariance statistics, the present formulation transforms modulation estimation into a localised multiscale operator problem. This transition is enabled by the localised freezing approximation derived in Section~\ref{sec4}. Wavelet coefficients become localised observations of multiplicative modulation structure whose covariance properties may be analysed scale-by-scale and position-by-position. 

Importantly, the resulting formulation is not exact; its validity depends explicitly on wavelet support geometry, local modulation regularity, scale-dependent support overlap, and residual multiscale mixing. 

\subsection{Classical covariance-based MRW unwrapping}
In multiplicatively modulated stochastic processes of the form
\begin{equation}
X(x)=e^{\omega(x)}\epsilon(x),
\end{equation}
\noindent
the hidden modulation field $\omega(x)$ controls the intermittent organisation of fluctuations across scales. In MRW-type models, the modulation process typically possesses logarithmic covariance structure of the form
\begin{equation}
\mathrm{Cov}[\omega(x),\omega(x+r)] \sim -\lambda^2\log r,
\end{equation}
over an appropriate scaling range.
Classical MRW unwrapping approaches exploit this logarithmic covariance structure operationally. In particular, covariance statistics computed from logarithmic amplitudes of multiscale observables may be used to estimate intermittency parameters and reconstruct aspects of the hidden modulation process.

In the original covariance-based formulations, multiscale observables were typically constructed using increments, scale-dependent averages, or Fourier-domain quantities. These formulations remain predominantly global in character. Covariance estimation is performed through large-scale averaging over the entire signal or field, while localisation enters only indirectly through scale dependence.

The present work develops a localised reformulation based on wavelet probing operators. Rather than estimating modulation statistics globally, localised wavelet coefficients are treated as finite-support observations whose geometry directly controls the validity of local multiplicative separation.

\subsection{Localised logarithmic wavelet formulation}
Let
\begin{equation}
W_X(a,b) = \int X(x)\psi_{a,b}(x)\,dx
\end{equation}
denote the continuous wavelet transform of the observed process
$X(x)$, where:
$a$ denotes scale, 
$b$ denotes position, 
and $\psi_{a,b}$ denotes the analyzing wavelet. 

Substituting the multiplicative representation
\begin{equation}
X(x)=e^{\omega(x)}\epsilon(x),
\end{equation}
yields
\begin{equation}
W_X(a,b) = \int e^{\omega(x)} \epsilon(x) \psi_{a,b}(x)\,dx.
\end{equation}
If the modulation field varies sufficiently slowly across the wavelet support, then
\begin{equation}
\omega(x)\approx\omega(b), \quad x\in\mathrm{supp}(\psi_{a,b}),
\end{equation}
leading to the localised freezing approximation
\begin{equation}
W_X(a,b) \approx e^{\omega(b)} W_\epsilon(a,b).
\end{equation}
Taking logarithmic amplitudes gives
\begin{equation}
\log \lvert W_X(a,b) \rvert \approx \omega(b) + \log |W_\epsilon(a,b)|. 
\label{eq:local_operator}
\end{equation}
\noindent
Equation Eq.~\eqref{eq:local_operator} constitutes the central localised operator relation underlying the present formulation.

Importantly, the approximation is intrinsically local. Its validity depends explicitly on the degree to which the modulation field remains approximately constant within the finite wavelet probing domain. The formulation therefore transforms modulation estimation into a localised covariance problem defined directly on wavelet coefficients.
Assuming weak correlation between the carrier contribution and the modulation process, the covariance of logarithmic wavelet amplitudes becomes approximately
\begin{equation}
\mathrm{Cov} [ \log \lvert W_X(a,b) \rvert, \log |W_X(a,b+r)| ] \approx \mathrm{Cov} [ \omega(b), \omega(b+r) ].
\end{equation}
Under MRW assumptions, this yields

\begin{equation}
\mathrm{Cov} [ \log \lvert W_X(a,b) \rvert, \log |W_X(a,b+r)| ] \sim -\lambda^2\log r.
\end{equation}
Thus, logarithmic covariance scaling may be estimated directly from localised wavelet amplitudes and this relation constitutes the localised wavelet analogue of classical MRW covariance unwrapping. Hence, the present viewpoint differs conceptually from classical multifractal wavelet approaches based on partition functions or wavelet-transform modulus maxima (WTMM) formalisms. In such approaches, locality primarily serves as a mechanism for characterising singularity organisation and estimating global multifractal statistics. Here, by contrast, localisation itself becomes the operational mechanism enabling approximate extraction and probing of the multiplicative modulation field. The wavelet representation is therefore not introduced merely as an alternative multiscale basis, but as a localised operator framework for MRW unwrapping.
Finally, it is important to distinguish between the conceptual and computational levels of the present formulation. The theoretical development is expressed using generic wavelet-transform notation
$W_X(a,b)$, emphasising localisation in both scale and position independently of discretisation. The numerical implementation discussed in subsequent sections employs a dyadic orthonormal discrete wavelet transform using compactly supported Daubechies wavelets. The continuous notation is adopted here because it provides the natural analytical language for describing local modulation freezing and multiscale localisation.

\subsection{Scale-dependent covariance estimation}
Unlike global covariance formulations, the localised wavelet framework depends fundamentally on support geometry. For a wavelet centered at $(a,b)$, the effective probing domain scales proportionally to the wavelet support width. Consequently, localisation quality, covariance estimation, and freezing validity all become explicitly scale dependent. 

This dependence produces a hierarchy of operational validity regimes. At fine scales, the support remains highly localised; modulation variability inside the support is weak; and localised factorisation remains approximately valid. Conversely, at coarse scales, support overlap increases; modulation variability across the support becomes substantial; and residual multiscale mixing emerges. This produces a hierarchy of operational validity regimes. As the analysis scale increases, the effective support of the wavelet expands (Figure~\ref{fig: wavelet_localisation}B), progressively increasing the probability of internal modulation variability and support overlap effects. The covariance distortions observed at large scales are therefore not anomalous effects but direct consequences of support-induced mixing. The geometric origin of this behaviour is already evident in the scale-dependent support expansion illustrated in Figure~\ref{fig: modulation_freezing}.

The covariance estimator used throughout the present work may be written schematically as
\begin{equation}
C_a(r) = \mathrm{Cov} [ \log \lvert W_X(a,b) \rvert, \log |W_X(a,b+r)| ].
\end{equation}
\noindent
The dependence on scale $a$ is essential. In the localised freezing regime {\color{black}(as illustrated in Figure~\ref{fig: modulation_freezing}A)}, one expects
\begin{equation}
C_a(r) \approx -\lambda^2\log r.
\end{equation}
Outside this regime, support-induced distortions progressively modify the covariance structure.

The localised formulation therefore naturally predicts scale-dependent estimator bias, finite-support distortions, and a progressive breakdown of logarithmic scaling at coarse scales. This behavior differs fundamentally from idealised asymptotic formulations of multiplicative modulation, in which scaling is often treated independently of localisation geometry.

\subsection{Residual support coupling and estimator bias} \label{sec: residual support}
The freezing approximation developed above is only approximately valid; residual modulation variability inside the wavelet support generates multiscale coupling terms neglected by the idealised factorisation. To make this explicit, we expand the modulation field as
\begin{equation}
    \omega(x) = \omega(b) + \delta\omega(x,b),
\end{equation}
where $\omega(b)$ denotes the locally frozen component and $\delta\omega(x,b)$ represents the residual modulation variability within the support. {\color{black} As illustrated schematically in Figure~\ref{fig: covariance_origin}B, the latter captures modulation variability occurring within the wavelet support itself.}
Substituting this expansion into the wavelet transform yields
\begin{equation}
    W_X(a,b) = e^{\omega(b)} \int e^{\delta\omega(x,b)} \epsilon(x)\psi_{a,b}(x)\,dx,
\end{equation}
where the residual field $\delta\omega(x,b)$ generates support-dependent multiscale coupling whose contribution grows progressively with scale. 

This mechanism has several important consequences: first, covariance estimation becomes biased by support-induced mixing terms; second, the effective covariance structure deviates progressively from ideal logarithmic scaling; and third, scale-dependent transition regions emerge in which localised freezing becomes only partially valid. Formally, this expansion allows the log-amplitude to be expressed as
\begin{equation}
    \log\lvert W_X(a,b) \rvert = \omega(b) + \log|W_\epsilon(a,b)| + M(a,b),
\end{equation}
where $M(a,b)$ denotes the effective multiscale mixing contribution generated by internal modulation variability. The resulting covariance therefore becomes
\begin{equation}
    \mathrm{Cov} \left[ \log\lvert W_X(a,b) \rvert, \log|W_X(a,b+r)| \right] = C_\omega(r) + C_{\mathrm{mix}}(a,r) + C_{\epsilon}(a,r),
\end{equation}
where $C_{\mathrm{mix}}(a,r)$ is the scale-dependent mixing-induced term.
{\color{black}The effective support-coupling mechanism represented by $M(a,b)$ corresponds schematically to the residual mixing process illustrated in in Figure~\ref{fig: covariance_origin}C.}

These effects are particularly important at intermediate and coarse scales where support overlap increases; modulation variability accumulates; and effective coupling extends across multiple local neighbourhoods simultaneously. The support dependence of multiscale mixing is especially important when compactly supported orthonormal wavelets are employed numerically. In such cases, the detailed geometry of the analysing wavelet, including oscillatory structure, vanishing moments, and effective autocorrelation properties, contributes directly to the shape of the mixing kernel. The resulting covariance distortions therefore depend not only on the modulation field statistics themselves, but also on the interaction between the modulation field and the multiscale geometry of the wavelet representation.

The numerical investigations presented in Section~\ref{sec6} strongly support this interpretation. In particular, the covariance “bump” structures observed experimentally appear consistent with support-induced residual multiscale mixing predicted by the present framework. Importantly, this interpretation transforms apparent deviations from ideal covariance scaling into theoretically meaningful finite-scale effects. Rather than indicating simple reconstruction failure, the observed distortions may instead provide direct information about effective support geometry, modulation variability, and localised multiscale coupling structure.

This observation suggests an important reinterpretation of the residual covariance structures encountered in practice. Rather than treating these deviations purely as numerical artifacts, they may instead be viewed as measurable signatures of finite-support multiscale coupling. From this perspective, the observed covariance residuals contain structural information regarding the interaction between local modulation field organisation and wavelet support geometry.
The present framework therefore leads naturally to the notion of an effective multiscale mixing kernel associated with localised modulation field probing. Such a kernel does not appear in classical global covariance formulations because it emerges specifically from localised finite-support observations of the modulation field. Understanding and characterising this effective mixing structure constitutes an important direction for future development of localised multifractal operator formulations.

Finally, residual multiscale mixing provides a direct explanation for the scale-dependent degradation of unwrapping performance observed numerically. At fine scales, local modulation freezing dominates and covariance estimates remain close to theoretical predictions. At intermediate and coarse scales, mixing progressively destroys the separability of modulation and carrier contributions, reducing the effective scaling range and ultimately limiting the validity of localised reconstruction. The transition from freezing-dominated to mixing-dominated behaviour therefore defines the operational validity range of wavelet-based MRW unwrapping.

To estimate the kernel the surrogate white noise was used. A number $n$ independent samples of white noise $\epsilon_i (x)$, $i = 1, \ldots n$, were generated and their wavelet transform was calculated using the same Daubechies wavelets, yielding the coefficients $W_{\epsilon, i} (a,b)$. For each realisation and scale $a$ the autocovariances $C_{\epsilon, i} (a, r)$ were calculated
\begin{equation}
    C_{\epsilon, i} (a, r) = \Cov \left[ \log \lvert W_{\epsilon, i} (a,b) \rvert, \log \lvert W_{\epsilon, i} (a,b+r) \rvert \right]
\end{equation}
and then averaged position-by-position to yield the estimation of noise covariance
\begin{equation}
    C_{\epsilon} (a,r) = \frac{1}{n} \sum_{i=1}^n C_{\epsilon, i} (a, r).
\end{equation}
To estimate the support extension we used as an underlying true process the wavelet-averaged $\omega$
\begin{equation}
    \omega_a(b) = \frac{\int \omega(x)\,|\psi_{a,b}(x)|\,dx} {\int |\psi_{a,b}(x)|\,dx}, \label{eq:omega_j}
\end{equation}
which in this case acts as a low-pass filter. Once again, $n$ samples of $\omega_i$ and respective $\omega_{a,i}$ were generated and their wavelet coefficients $W_{\omega, i} (a,b)$ and $W_{\omega_a, i} (a,b)$ calculated. These coefficients were then used to calculate averaged covariances in the similar manner as before, namely
\begin{equation}
    C_{\mathrm{full}} (a,r) = \frac{1}{n} \sum_{i=1}^n\Cov \left[ \log \lvert W_{\omega, i} (a,b) \rvert, \log \lvert W_{\omega, i} (a,b+r) \rvert \right],
\end{equation}
and similarly for $\omega_a$ and its log-coefficient covariance $C_{\omega_a} (a,r)$. The mixing term was then estimated by subtracting the smooth and noise terms from the full covariance
\begin{equation}
    C_{\mathrm{mix}} (a,r) = C_{\mathrm{full}} (a,r) - C_{\omega_a} (a,r) - C_{\epsilon} (a,r). \label{eq:mix cov}
\end{equation}

\subsection{Practical wavelet implementation considerations}
Although the preceding derivations were formulated using the continuous wavelet transform for analytical clarity, the numerical implementation developed in the present work employs discrete orthonormal wavelet decompositions. This distinction is conceptually important. The continuous formulation provides the natural analytical framework because localisation, support geometry, and multiscale probing are most transparently expressed through continuous wavelet operators. The discrete implementation should therefore be interpreted primarily as a numerical discretisation of the underlying localised operator framework rather than as a fundamentally distinct formulation.

In practice, compactly supported orthonormal Daubechies wavelets were employed in order to maintain strong localisation, preserve multiscale structure, and control support overlap.

Daubechies wavelets are uniquely characterised by possessing the maximal number of vanishing moments for any given support length. They are denoted $\mathrm{db}N$, where $N$ indicates the number of vanishing moments, e.g. for $\mathrm{db}4$ the first four moments vanish. Furthermore, for a given $N$ they have the compact support of length $2N-1$ and any other wavelet family with $N$ vanishing moments would have a larger support length~\cite{daubechies88, daubechies92}. In the discrete setting the scale parameter $a$ is replaced by $j$, which dictates the dyadic dilation (where $a = 2^j$), while the continuous translation parameter $b$ is replaced by an integer $k$ representing the discrete shift. Hence the continuous definition given in eq.~\eqref{eq:cwt wavelet definition} is replaced by
\begin{equation}
    \psi_{j,k} (x) = \frac{1}{2^{jd/2}} \psi\left( \frac{x-2^j k}{2^j} \right).
\end{equation}
Throughout the rest of this paper, we shall replace the parameters $a$ and $b$ with $j$ and $k$ respectively, denoting the finest scale (mother wavelet) as $j=0$, where the support width of the wavelets increases with increasing $j$. Wavelet support width plays a central role because it directly determines probing-domain geometry, freezing validity, and residual multiscale mixing, while additional practical considerations include boundary effects, finite sample size, covariance averaging procedures, and scale-dependent coefficient statistics.

\subsection{Operational validity of localised freezing}
The localised wavelet formulation developed here does not define a universally valid decomposition. Instead, it defines a scale-dependent operational approximation whose validity depends on the relation between wavelet support geometry and local modulation variability. The framework therefore naturally leads to adaptive validity regimes. At sufficiently fine scales, localisation dominates; local modulation freezing becomes approximately valid; and covariance estimation remains reliable. At intermediate scales, residual support coupling emerges progressively; covariance distortions appear; and estimator bias increases. Finally, at sufficiently coarse scales, support-induced multiscale mixing dominates; localised separation breaks down; and identifiable logarithmic scaling ranges collapse.

This viewpoint is conceptually important because it reframes localised MRW unwrapping as a geometrically controlled multiscale operator problem rather than a purely statistical estimation procedure. The numerical investigations presented in the following section demonstrate that many experimentally observed covariance deviations are quantitatively consistent with the support-controlled finite-scale breakdown predicted by the present framework. From this perspective, localised wavelet operators provide not only multiscale representations but operational access to local multiplicative organisation itself.

\section{Numerical Validation and Finite-Scale Mixing}\label{sec6}
The theoretical developments presented in Sections~\ref{sec2},~\ref{sec3} and~\ref{sec4} suggest that localised wavelet probing may induce approximate multiplicative separation over finite-support observation domains. The purpose of the present section is to investigate this hypothesis numerically using synthetic multifractal random walk (MRW) realisations together with localised wavelet decompositions.

The numerical investigations pursue four primary objectives: (i)~validation of localised logarithmic covariance scaling predicted by the freezing approximation; (ii)~characterisation of residual covariance distortions generated by finite-support multiscale mixing; (iii)~investigation of effective support corrections and surrogate kernel estimation; and (iv)~evaluation of scale-dependent reconstruction quality and freezing validity. Particular emphasis is placed on understanding the transition between localised freezing regimes, intermediate multiscale mixing regimes, and coarse-scale breakdown regimes. 

Importantly, the numerical experiments are not intended merely as empirical demonstrations of a reconstruction procedure. Rather, they serve as structured tests of the localised freezing framework itself.

\subsection{Synthetic MRW generation and wavelet implementation}
Synthetic MRW realisations were generated using logarithmically correlated Gaussian modulation fields together with multiplicative stochastic carrier processes. The simulations were designed to reproduce the covariance structure characteristic of intermittent multiplicative modulation models while allowing direct access to the underlying hidden modulation field
$\omega(x)$.
The observed field was generated according to
\begin{equation}
X(x)=e^{\omega(x)}\epsilon(x),
\end{equation}
where:
$\omega(x)$ denotes the modulation field, 
and $\epsilon(x)$ denotes the carrier fluctuation process. 

In the practical MRW generation the procedure described by Lakhal \textit{et al.} in~\cite{Lakhal25} was used. The multifractal process can be constructed by fractionally integrating the non-linear transformation of log-correlated Gaussian field. Let $\omega (x)$ be a log-correlated Gaussian field with cutoff $\xi$, i.e. it's correlation function is given by
\begin{equation}
    C_{\omega} (r) = \lambda^2 K_0 ( r / \xi ),
\end{equation} 
where $K_0$ is the modified Bessel function of the second kind with parameter 0. For $r \ll \xi$ this correlation can be approximated by $C_{\omega} (r) \approx - \lambda^2 \ln (r / \xi)$. The log-correlated field can be obtained by fractionally integrating the white noise
\begin{equation}
    \omega (x) = \lambda \left( \xi^{-2} - \Delta \right)^{-d/4} \eta (x),
\end{equation}
where $\eta$ is a $d$-dimensional white noise and $(-\Delta)^{\alpha/2}$ is a fractional Laplacian operator with exponent $\alpha$. In practice, the fractional integration is carried in the Fourier domain, in which the fractional Laplacian acts as a filter $G ( \boldsymbol{k}) = 1/\boldsymbol{k}^{\alpha}$. In the lattice lattice of step one $\lVert \boldsymbol{k} \rVert^2$ is replaced by the $2 \times \sum_{i=1}^d \left(1 - \cos 2\pi k_i \right)$.

We restricted our focus to one-dimensional MRW processes for clarity, although the procedure can be used in higher dimensions as well. In this setting we have found that the intermittency parameter has to be sufficiently large for the model to exhibit multifractal properties. More precisely, we require that $\lambda^2$ is of the order of the length of the process $N$. Otherwise, if $\lambda$ is too small, the behaviour process $X(x) = \epsilon(x) e^{\omega(x)}$ is dominated by the noise $\epsilon$. MRW process samples has a length of $N = 2^{14}$.

Wavelet decompositions were performed using compactly supported orthonormal Daubechies wavelets implemented through stationary wavelet transforms (SWT). The use of compact support is particularly important because the localised freezing approximation depends explicitly on finite-support probing geometry. The baseline wavelet family consists of Daubechies $\mathrm{db}4$ wavelets possessing four vanishing moments, where the mother wavelet has a support length of seven~\cite{daubechies88}. Before computing the SWT, the fields were padded symmetrically at the boundaries to ensure proper edge handling. The numerical experiments investigate both idealised synthetic validation scenarios and progressively more difficult regimes involving coarse-scale support overlap and residual multiscale coupling.

\subsection{Validation of localised logarithmic covariance scaling}
The first objective of the numerical analysis is to test whether localised wavelet coefficients reproduce the logarithmic covariance structure predicted by the freezing approximation. Under localised multiplicative separation,

\begin{equation}
\log \lvert W_X(a,b) \rvert \approx \omega(b) + \log |W_\epsilon(a,b)|,
\end{equation}
\noindent
the covariance of logarithmic wavelet amplitudes is expected to inherit the covariance structure of the underlying modulation field. In particular, one expects approximate logarithmic covariance scaling over scales for which local freezing remains valid.

\begin{figure}[htb]
    \centering
    \includegraphics[width=\linewidth]{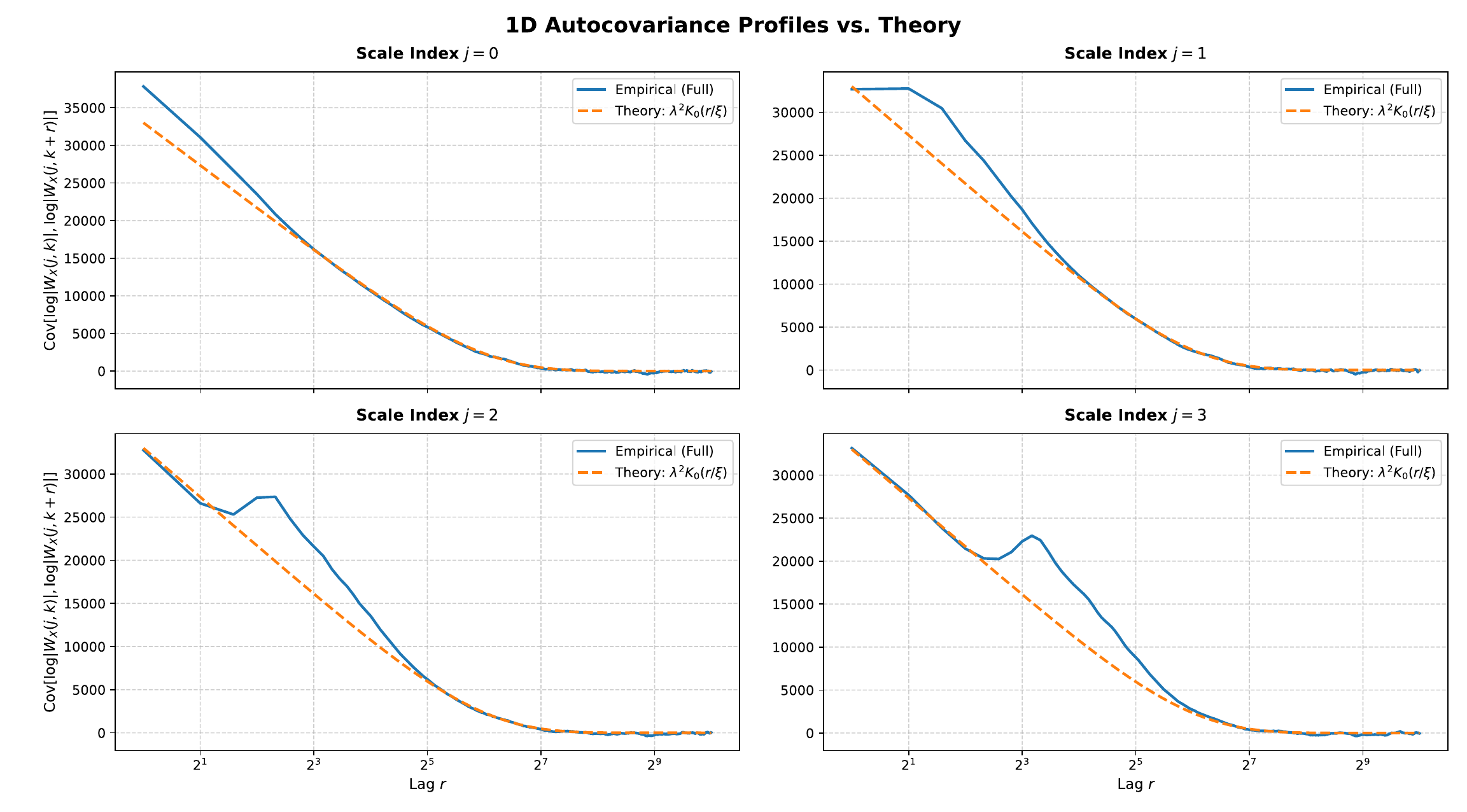}
%    \includesvg[width=\linewidth]{figure7.svg}
    \caption{Autocovariance plots of the logarithmic wavelet coefficiets of a one-dimensional intermittent field. Empirical covariance is plotted with a solid line, while theoretical prediction is indicated by a dashed line. Plotted for parameters $N=2^{14}$, $\lambda^2= 0.5 \times N$, $\xi = 50$ and $n=100$ with $\mathrm{db}4$ wavelets used.} \label{fig: 1d_cov}
\end{figure}

We are primarily focused on the autocovariance function of the logarithmic wavelet coefficients calculated as
\begin{equation}
    C (j, r) = \Cov \left[ \log |W_X(j,k)|, \log |W_X(j,k+r)| \right].
\end{equation}
Figure~\ref{fig: 1d_cov} presents the covariance estimates obtained from wavelet coefficients across multiple scales.

At fine scales, the numerical results demonstrate substantial agreement between empirical covariance estimates and the theoretical logarithmic scaling predicted by the modulation model. In particular, the covariance slope remains approximately consistent with the prescribed intermittency parameter over a significant intermediate lag range. This behavior strongly supports the central hypothesis that sufficiently localised wavelet probing domains induce approximate local multiplicative separation. The agreement is especially strong at fine scales where the wavelet support remains highly localised; modulation variability across the support remains limited; and residual multiscale mixing remains weak. The numerical results therefore support the interpretation that wavelet localisation creates effective local probing domains over which modulation freezing becomes approximately valid.

At progressively coarser scales, however, systematic deviations begin to emerge. These effects become particularly important in the intermediate-lag regime and are investigated in detail below.

\subsection{Emergence of residual multiscale mixing}
\begin{figure}[htb]
    \centering
    \includegraphics[width=\linewidth]{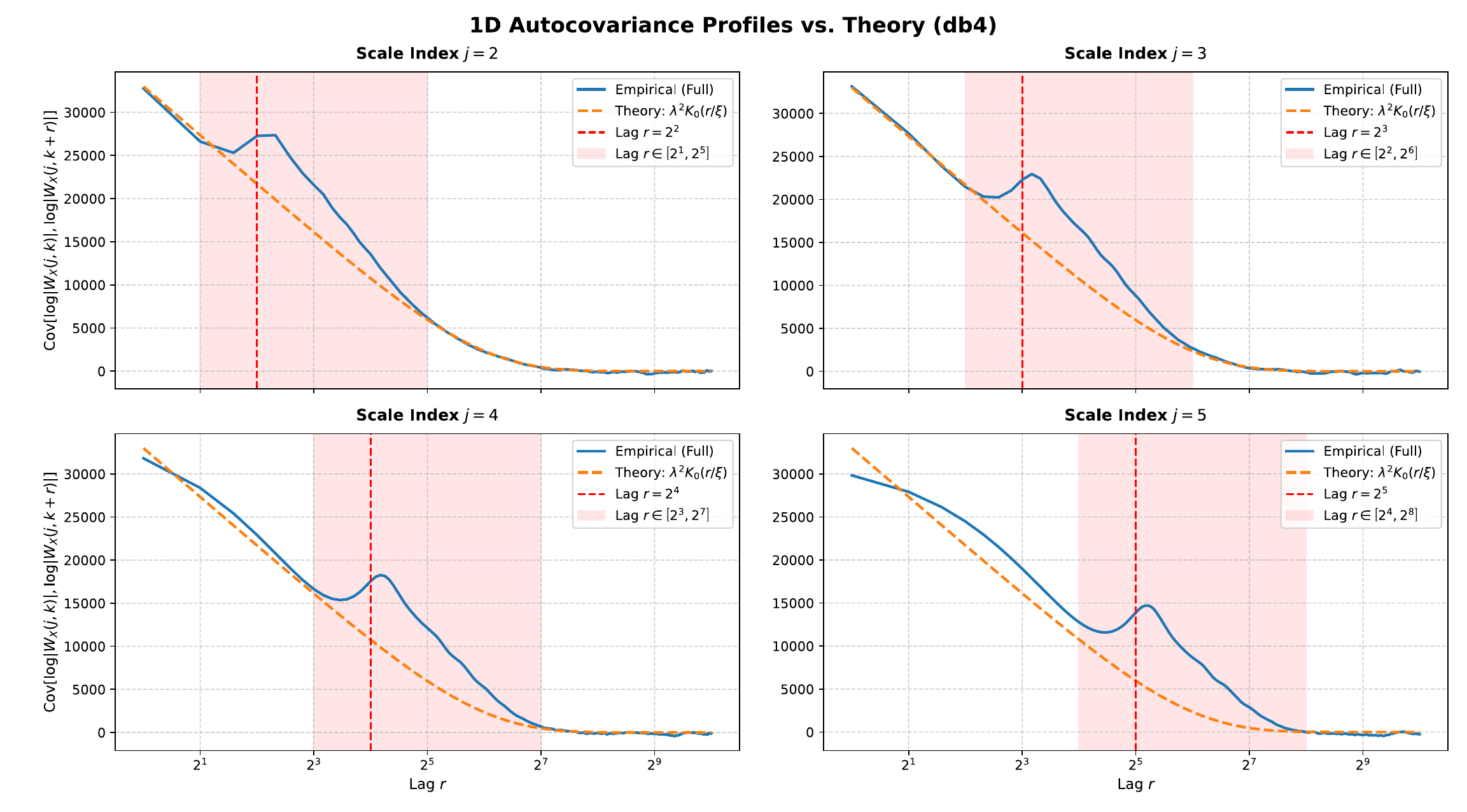}
%    \includesvg[width=\linewidth]{figure8.svg}
    \caption{Bump formation in the intermediate scales. For a given scale $j$ the bumps are located around the lag $r = 2^{j}$, which corresponds to the scaling parameter of a wavelet at that resolution. The bump width is approximately the length of wavelet support, i.e. the length of mother wavelet support (in case of $\mathrm{db}4$ wavelets this is equal to $L = 7$) times the scaling factor $2^j$. Plotted for parameters $N=2^{14}$, $\lambda^2= 0.5 \times N$, $\xi = 50$ and $n=100$ with $\mathrm{db}4$ wavelets used.} \label{fig: 1d_bump}
\end{figure}

\begin{figure}[htb]
    \centering
    \includegraphics[width=\linewidth]{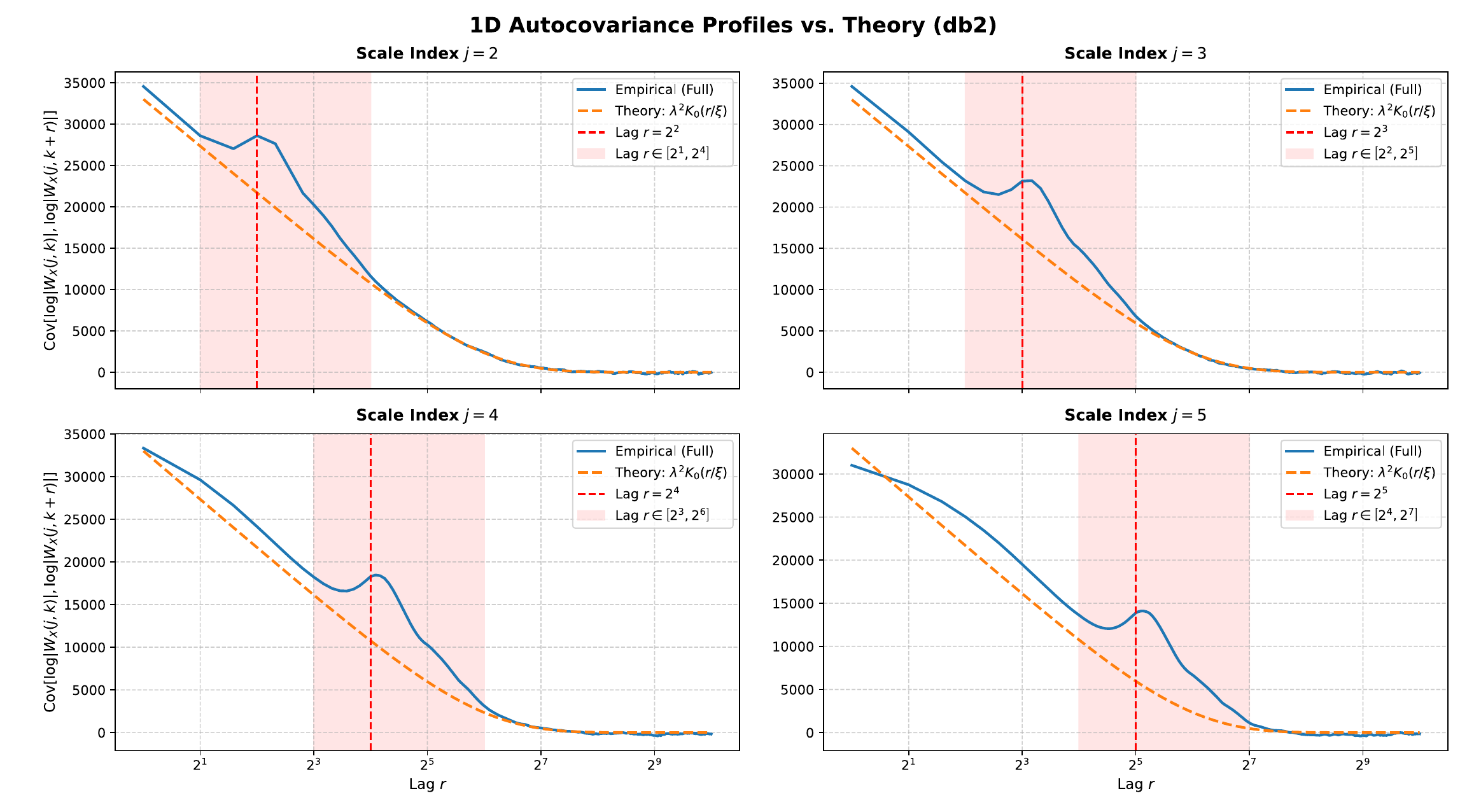}
%    \includesvg[width=\linewidth]{figure9.svg}
    \caption{Bump formation in the intermediate scales. For a given scale $j$ the bumps are located around the lag $r = 2^{j}$, which corresponds to the scaling parameter of a wavelet at that resolution. The bump width is approximately the length of wavelet support, i.e. the length of mother wavelet support (in case of $\mathrm{db}2$ wavelets this is equal to $L = 3$) times the scaling factor $2^j$. Plotted for parameters $N=2^{14}$, $\lambda^2= 0.5 \times N$, $\xi = 50$ and $n=100$ with $\mathrm{db}2$ wavelets used.} \label{fig: 1d_bump_db2}
\end{figure}

One of the most significant observations emerging from the numerical experiments is the appearance of systematic covariance distortions at intermediate and coarse scales. Representative examples are shown in Figure~\ref{fig: 1d_bump} for $\mathrm{db}4$ wavelets and Figure~\ref{fig: 1d_bump_db2} for $\mathrm{db}2$ wavelets.

Several characteristic features emerge consistently: fine scales exhibit near-ideal logarithmic covariance scaling; intermediate scales develop secondary covariance structures; and coarse scales display a progressive breakdown of identifiable scaling ranges. Most notably, covariance estimates exhibit localised “bump”-like distortions whose position and extent evolve systematically with wavelet scale. Initially, such deviations might naturally be interpreted as numerical artifacts, reconstruction instability, imperfect simulation, or surrogate-model failure. However, the theoretical developments of Section~\ref{sec4} suggest a different interpretation. As wavelet support increases with scale, the probing domain progressively incorporates heterogeneous modulation neighbourhoods simultaneously. The modulation field can no longer be approximated as locally frozen over the entire support region. Consequently, the wavelet coefficient contains coupled contributions arising from multiple local modulation structures, a mechanism that generates residual multiscale mixing.

The numerical results strongly support this interpretation. In particular, the extent of covariance distortion grows systematically with support size; fine scales remain comparatively unaffected; and residual coupling appears concentrated precisely within the transition regime predicted by support-controlled freezing breakdown. For lags smaller than the wavelet support ($r < 2^j$), the internal ~{\color{black}spatial} mixing heavily smooths the field, pulling the empirical covariance below the theoretical curve as the sharp local fluctuations are averaged out. At the resonant lag ($r \approx 2^j$), the convolution creates an artificial correlation peak, pushing the empirical line above the theory. As the scale increases, the bump moves toward larger lags. More precisely, for a given scale $j$, the bump maximum is located around the lag $r=2^j$, which corresponds to the scaling parameter of a wavelet at that resolution. 

The width of the bump is roughly equal to the support length of the wavelet at that scale. For a $\mathrm{db}N$ wavelet at scale $j$, the support length is given by $L_j = L_0 2^j$, where $L_0 = 2N-1$ is the support length of the mother wavelet (e.g., $L_0 = 7$ for $\mathrm{db}4$ and $L_0 = 3$ for $\mathrm{db}2$). By comparing Figures~\ref{fig: 1d_bump} and~\ref{fig: 1d_bump_db2}, one can observe that the bump width is twice as large when $\mathrm{db}4$ wavelets are employed. The emergence of residual covariance structure therefore appears to constitute a physically meaningful signature of finite-support multiscale mixing rather than merely a numerical imperfection. An especially important observation is that the covariance distortions are not random; instead, they exhibit reproducible, scale-dependent organisation suggestive of effective support interactions generated within the localised probing domain itself.

\subsection{Effective support extension and surrogate kernel estimation}
\begin{figure}[htb]
    \centering
    \includegraphics[width=\linewidth]{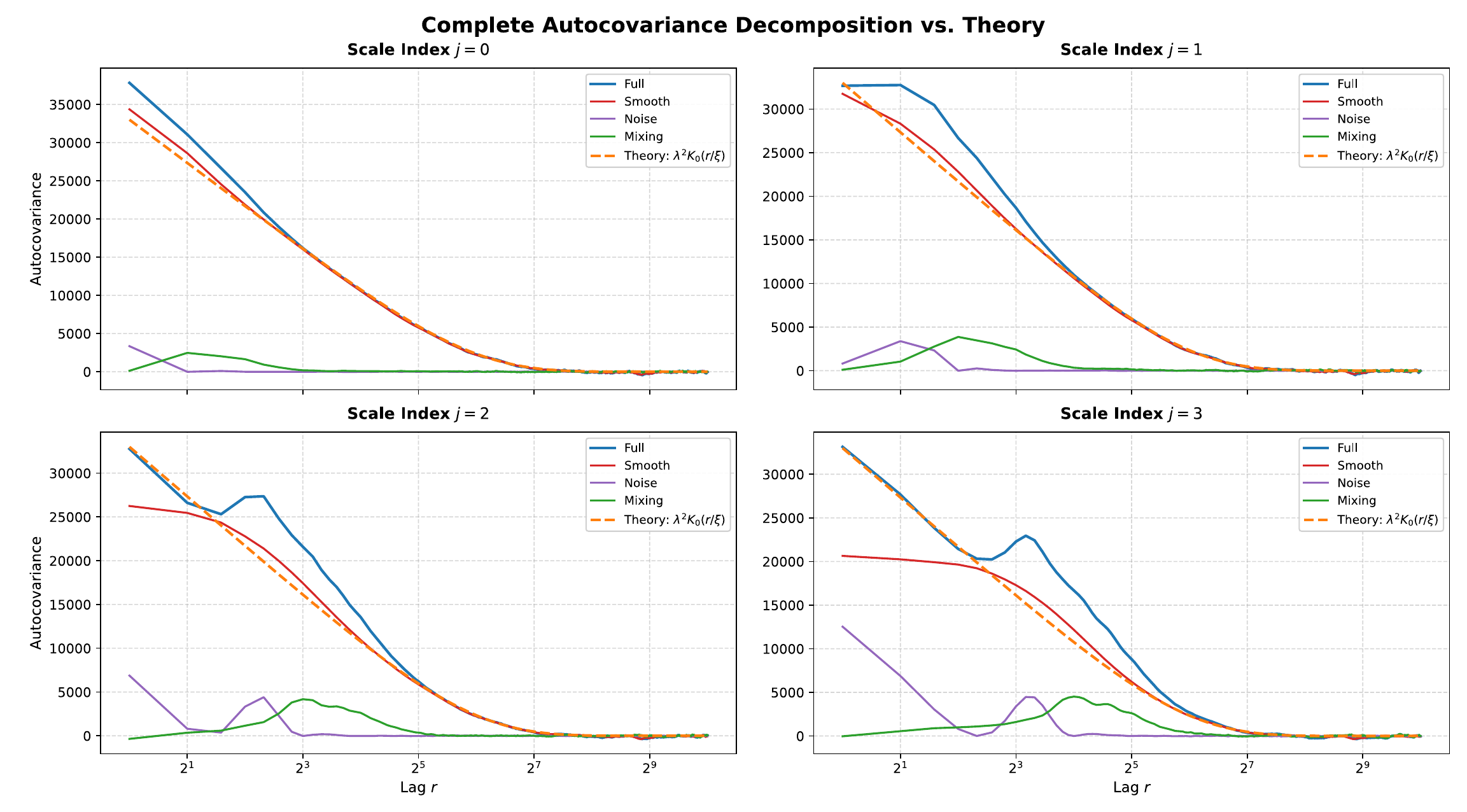}
%    \includesvg[width=\linewidth]{figure10.svg}
    \caption{Surrogate kernel estimation. The full empirical covariance $C_{\mathrm{full}}$ is plotted in blue, and that of the smoothed process $C_{\omega_j}$ is shown in red. The kernel covariance, estimated according to Section~\ref{sec: residual support}, is shown in purple, while the mixing covariance obtained through \eqref{eq:mix cov} is plotted in green. The kernel covariance is responsible for the bump maximum at $r = 2^j$, whereas the mixing covariance widens the bump. Plotted for parameters $N=2^{14}$, $\lambda^2= 0.5 \times N$, $\xi = 50$, and $n=100$, utilising $\mathrm{db}4$ wavelets.} \label{fig: 1d_kernel_cov}
\end{figure}

In order to investigate the origin of the residual covariance distortions further, surrogate support-kernel estimation experiments were performed. The central idea is to estimate the covariance contribution generated purely by wavelet support geometry and subtract this contribution from the observed covariance estimates. Such surrogate corrections attempt to isolate the idealised frozen component from residual support-induced coupling. Figure~\ref{fig: 1d_kernel_cov} presents representative examples of raw covariance estimates, estimated support kernels, and mixing covariance estimates obtained from \eqref{eq:mix cov}.

The surrogate corrections successfully reproduce several important features of the observed covariance distortions, including the principal intermediate-scale bump structure, support-dependent covariance broadening, and scale-dependent transition behavior. However, residual discrepancies remain even after subtraction of the estimated support contribution, which constitutes an important observation. As discussed in the previous section, the bump maximum is located around the lag $r = 2^{j}$ and the width of the bump is roughly equal to the support length of the wavelet at the given scale. Figure~\ref{fig: 1d_kernel_cov} clearly illustrates the individual components that constitute the bump. The covariance distortions associated with the surrogate white noise kernel are located near $r = 2^j$ and possess a width of approximately $2^{j+1}$; these are responsible for the observed bump maximum. Furthermore, the observed increase in covariance as $r \to 1$ at intermediate scales compensates for the ~{\color{black}spatial} smoothing of the field. Since the white noise is ~{\color{black}spatial}ly uncorrelated, this part of the bump is a direct result of the geometry of the wavelet itself, leaving the mixing covariance $C_{\mathrm{mix}}$ as the remaining contribution.

These remaining deviations suggest that the effective interaction structure may extend beyond the nominal support geometry of the analyzing wavelet itself. In other words, the multiscale coupling process may generate effective support extensions associated with modulation-induced interaction between neighbouring support regions. Although preliminary, this interpretation is conceptually significant because it suggests that localised multiscale operators may develop emergent interaction scales exceeding their nominal analytical support. At present, the precise mathematical structure of these effective support extensions remains unresolved. 

Several possible mechanisms may contribute, including residual modulation variability within the support, wavelet autocorrelation structure, support overlap interactions, finite realisation effects, and local regularity transitions. Further theoretical investigation will be required in order to characterise these effects rigorously. Nevertheless, the numerical results strongly suggest that residual covariance distortions are structured and reproducible consequences of finite-support multiscale coupling rather than arbitrary numerical noise.

\subsection{Scale-dependent reconstruction quality}

\begin{figure}[htb]
    \centering
    \includegraphics[width=\linewidth]{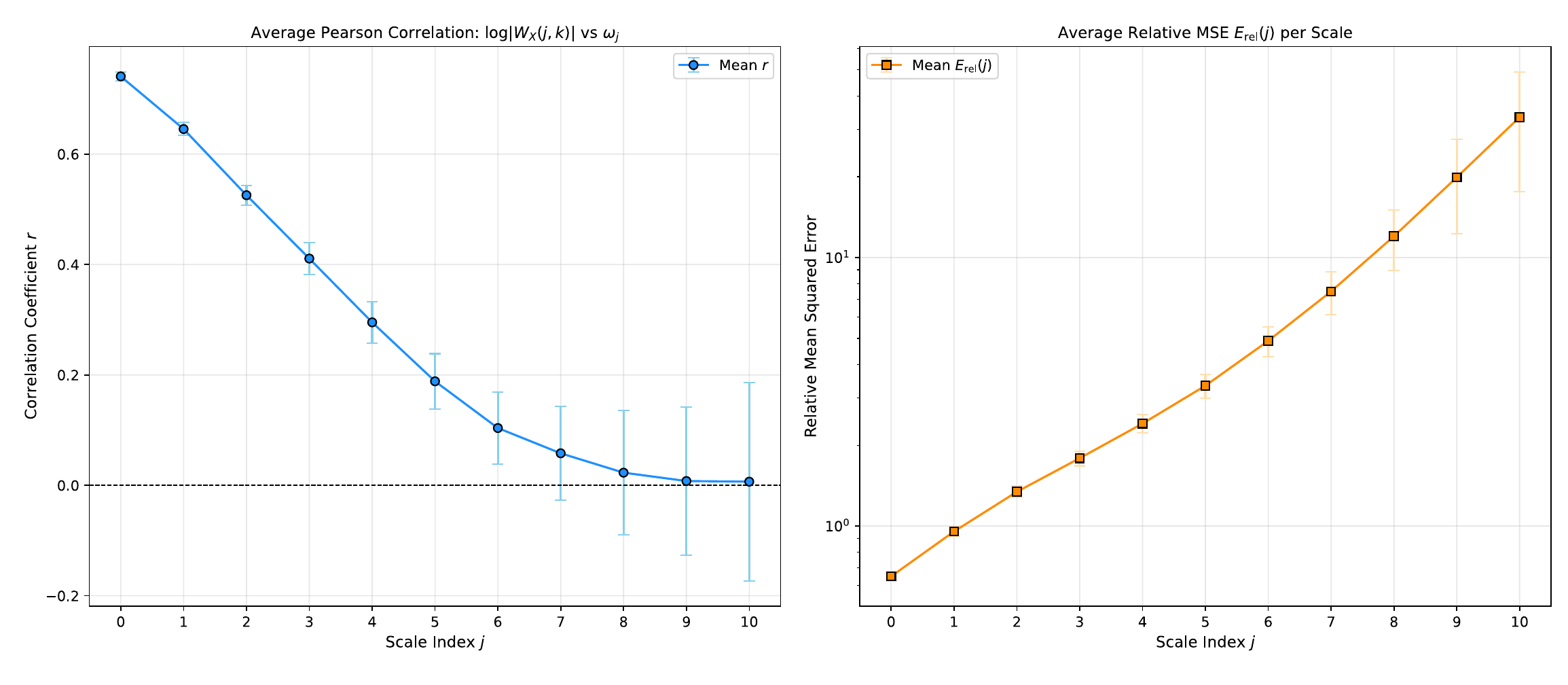}
%    \includesvg[width=\linewidth]{figure11.svg}
    \caption{Pearson correlation coefficient $r$ (left) and relative mean square error 
    (right) of logarithmic wavelet coefficients $\log \lvert W_X (j,k) \rvert$ and 
    the averaged process $\omega_j$ per wavelet scale $j$. Relative MSE per level is 
    defined in \eqref{eq: relative level mse}. Plotted for parameters $N=2^{14}$, 
    $\lambda^2= 0.5 \times N$, $\xi = 50$, and $n=100$, utilising $\mathrm{db}4$ wavelets.} 
    \label{fig: pearson_error}
\end{figure}

The freezing framework developed in this work naturally predicts that reconstruction quality should depend strongly on scale. To investigate this behavior quantitatively, reconstruction quality was evaluated using scale-dependent Pearson correlation, normalised reconstruction error, and energy-weighted reconstruction metrics. Crucially, the reconstruction does not recover the point-wise modulation field, but rather its scale-dependent coarse-grained representation consistent with the support of the analysing wavelet; to assess the reconstruction quality, we compare the reconstructed field with $\omega_j$ defined in \eqref{eq:omega_j}. Figure~\ref{fig: pearson_error} presents representative reconstruction statistics across scales.

The results reveal a systematic transition between strong fine-scale agreement, intermediate-scale degradation, and coarse-scale breakdown. In particular, the Pearson correlation between the reconstructed and ground-truth modulation fields increases progressively toward finer scales, while the normalised reconstruction error displays strong scale dependence. Although these results may initially appear counterintuitive, within the localised freezing framework they admit a natural interpretation. Fine-scale coefficients remain highly localised but also contain strong carrier variability and limited support averaging. Coarser scales provide greater averaging stability but simultaneously suffer from increased support-induced multiscale mixing. 

The reconstruction quality therefore reflects a competition between localisation fidelity, carrier-noise suppression, and support-induced modulation coupling. This interpretation is consistent with the progressive covariance distortions observed in the previous subsections. To account for the strongly nonuniform energy distribution across scales, energy-weighted reconstruction metrics were additionally considered. These metrics provide an improved characterisation of reconstruction quality by emphasising scales carrying the dominant wavelet energy contribution.

\begin{figure}[htb]
    \centering
    \includegraphics[width=\linewidth]{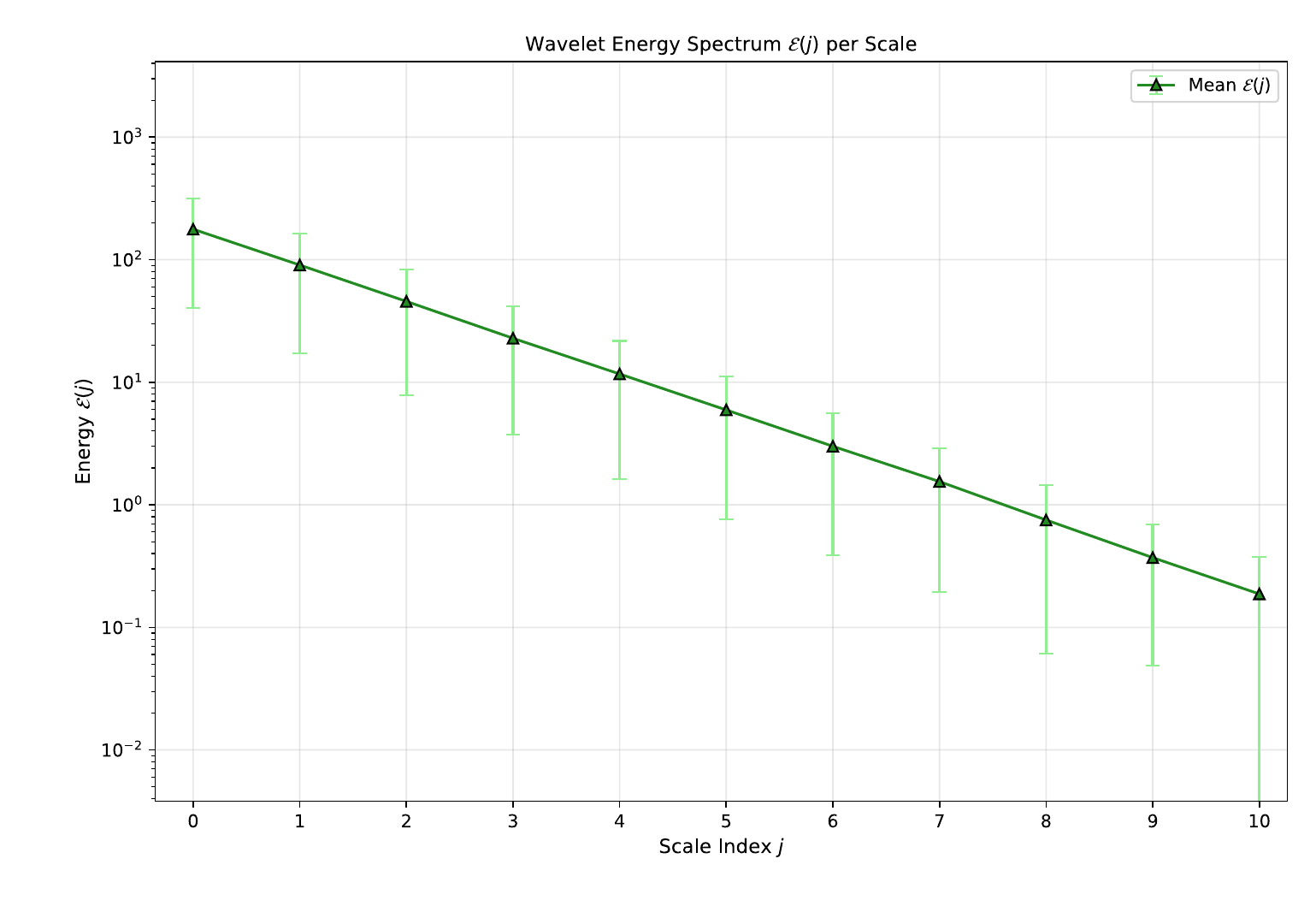}
%    \includesvg[width=\linewidth]{figure12.svg}
    \caption{Mean wavelet energy per scale defined as per eq.~\eqref{eq: wavelet energy}. Plotted for parameters $N=2^{14}$, $\lambda^2= 0.5 \times N$, $\xi = 50$ and $n=100$ with $\mathrm{db}4$ wavelets used.} \label{fig: wavelet_energy}
\end{figure}

Let us define the per scale wavelet energy as
\begin{equation}
    \mathcal{E} (j) = \frac{1}{N} \sum_{k} \lvert W_X (j,k) \rvert^2. \label{eq: wavelet energy}
\end{equation}
and per scale mean square error as
\begin{equation}
    E(j) = \frac{1}{N} \sum_{k} \left( \hat{\omega}_j (k) - \omega_j (k) \right)^2.
\end{equation}
To make this error relative per scale we take
\begin{equation}
    E_{\textrm{rel}} (j) = \frac{\frac{1}{N} \sum_{k} \left( \hat{\omega}_j (k) - \omega_j (k) \right)^2}{\frac{1}{N} \sum_{k} \left( {\omega}_j (k) - \bar{\omega}_j \right)^2}. \label{eq: relative level mse}
\end{equation}
Finally, the total energy-weighted relative error is given by
\begin{equation}
    E_{\textrm{rel}} = \frac{\sum_{j} \mathcal{E} (j) E_{\textrm{rel}} (j)}{\sum_{j} \mathcal{E} (j)}. \label{eq: relative mse}
\end{equation}
Since the small scales more precisely capture the local fluctuations they better reconstruct the underlying field $\omega$. At the same time, finer scales have much higher mean energy, as shown in figure~\ref{fig: wavelet_energy}. Hence weighting the error by level energy produces a single scalar performance metrics, while emphasising the most important scales.

Table~\ref{tab: performance} summarises reconstruction performance for a given set of parameters $\lambda^2$ and $\xi$. The Pearson correlation increases significantly with $\lambda^2$, while the relative MSE decreases accordingly. Both measures improve with increasing $\xi$ as well. A larger $\xi$ means the structural correlations stretch further across the~{\color{black} spatial} domain. When the field is more broadly correlated, there is less volatile internal modulation variability inside the wavelet support window. This directly reduces the residual support mixing and covariance distortions modeled in the theoretical framework. Higher values of $\lambda^2$ increase the amplitude and dominance of the log-correlated modulation structures over background fluctuations. The signal becomes more distinctly structured, allowing the wavelet coefficients to lock onto the true underlying modulation field $\omega$ more accurately.

\begin{table}[h]
    \centering
    \caption{Performance metrics for reconstruction of the log-correlated field $\omega$: Pearson correlation coefficient $r$ for the finest scale $j=0$ and energy-weighted relative MSE (defined in~\eqref{eq: relative mse}) between logarithmic wavelet coefficients $\log |W_X (j,k)|$ and $\omega_j$ calculated for different sets of parameters. Length of signal used $N=2^{14}$ and number of realisations used $n=100$.} \label{tab: performance}
    \begin{tabular}{|c|c|c|c|}
    \hline
    \multicolumn{2}{|c|}{\textbf{Parameters}} & \textbf{Pearson $r$ for level $j=0$} & \textbf{Energy-weighted Relative MSE} \\ \hline

    % Rows 2 & 3: First lambda group, alternating xi
    \multirow{2}{*}{$\lambda^2 = 0.25 \cdot N$} & $\xi = 50$  &  $0.6544 \pm 0.009$  &  $1.76 \pm 0.11$  \\
                                            & $\xi = 100$ & $0.6827 \pm 0.0105$  & $1.4465 \pm 0.12$  \\ \hline

    % Rows 4 & 5: Second lambda group, alternating xi
    \multirow{2}{*}{$\lambda^2 = 0.5 \cdot N$}  & $\xi = 50$  & $0.7430 \pm 0.0075$  & $1.084 \pm 0.084$  \\
                                            & $\xi = 100$ & $0.7660 \pm 0.0084$  & $0.912 \pm 0.083$ \\ \hline

    % Rows 6 & 7: Third lambda group, alternating xi
    \multirow{2}{*}{$\lambda^2 = 0.75 \cdot N$} & $\xi = 50$  & $0.7773 \pm 0.0068$  & $0.876 \pm 0.066$ \\
                                            & $\xi = 100$ & $0.7994 \pm 0.0074$ & $0.726 \pm 0.071$ \\ \hline

    % Rows 8 & 9: Fourth lambda group, alternating xi
    \multirow{2}{*}{$\lambda^2 = 1 \cdot N$}    & $\xi = 50$  & $0.7952 \pm 0.0069$ & $0.775 \pm 0.084$ \\
                                            & $\xi = 100$ & $0.8168 \pm 0.0076$ & $0.645 \pm 0.067$\\ \hline
    \end{tabular}
\end{table}

\section{Discussion}\label{Discussion}
The present work develops a localised wavelet formulation of multiplicative modulation analysis motivated by multifractal random walk (MRW) unwrapping and related inverse multiscale problems. The central idea underlying the proposed framework is that wavelet localisation may act not merely as a representational property of multiscale decompositions, but as an operational mechanism capable of inducing approximate local multiplicative separation. More precisely, the manuscript demonstrates that localised wavelet probing naturally leads to approximate relations of the form
\begin{equation}
W_X(a,b) \approx e^{\omega(b)} W_\epsilon(a,b),
\end{equation}
or equivalently,
\begin{equation}
\log \lvert W_X(a,b) \rvert \approx \omega(b) + \log |W_\epsilon(a,b)|.
\end{equation}
These relations emerge because finite-support wavelets probe the modulation field only over localised neighbourhoods whose ~{\color{black}spatial} extent is explicitly controlled by scale. The resulting approximation, referred to throughout the manuscript as local modulation freezing, provides a geometrical and operator-based interpretation of localised MRW unwrapping. Importantly, the validity of the approximation is not universal; instead, it depends explicitly on wavelet support geometry, the local regularity of the modulation field, scale-dependent support overlap, and residual multiscale coupling generated within finite observation domains. This viewpoint leads naturally to a reinterpretation of several numerical effects observed throughout the manuscript.

\subsection{From global multifractal statistics to localised multiscale probing}
Classical multifractal analysis has historically focused on global statistical characterisation. Multiplicative cascades, MRW models, partition-function methods, and wavelet multifractal formalisms typically estimate scaling exponents, multifractal spectra, covariance laws, or global scaling functions. Even in wavelet-based multifractal approaches such as the WTMM framework, localisation primarily served as a mechanism for identifying singular structures whose statistics were subsequently aggregated globally. 

The present work adopts a different perspective. Rather than using localisation solely as an intermediate step toward global statistics, localisation itself is treated as the primary operational mechanism underlying modulation extraction. In this formulation, wavelets define finite-support multiscale probing domains over which multiplicative modulation may become approximately separable from the carrier fluctuations. This distinction is conceptually important.

The transition proposed here may be summarised schematically as a shift from global statistical characterisation to localised multiscale probing:
\begin{equation}
    \text{Global Statistical Characterisation} \quad \longrightarrow \quad \text{Localised Multiscale Probing}.
\end{equation}
Within this framework, wavelet coefficients are interpreted not simply as multiscale descriptors but as localised observations of modulation structure itself. This perspective may help bridge several historically related but only partially connected areas, including multifractal stochastic modelling, wavelet singularity analysis, inverse multiscale problems, localised intermittency analysis, and operator-based regularity theory. The formulation also suggests that many covariance-based multifractal procedures may admit localised operator interpretations beyond the specific MRW setting considered here.

\subsection{Local modulation freezing and support-controlled validity}
The freezing approximation developed in Section~\ref{sec4} constitutes the central theoretical mechanism of the manuscript. Its validity depends fundamentally on the relation between modulation variability and wavelet support geometry. At sufficiently fine scales, the support of the analyzing wavelet becomes small relative to the ~{\color{black}spatial} variability of the modulation field. In this regime, the modulation process varies only weakly across the support, leading to an approximate local factorisation. Conceptually, the wavelet transform creates localised observation domains within which the modulation field appears approximately frozen. 

This interpretation introduces a fundamentally geometrical viewpoint on multiplicative modulation extraction. The quality of local factorisation is determined not only by the stochastic properties of the modulation process itself, but also by support size, support overlap, wavelet regularity, and local smoothness properties of the modulation field. The freezing approximation therefore defines a scale-dependent validity regime rather than an asymptotically exact universal decomposition. The numerical investigations presented in Section~\ref{sec6} strongly support this interpretation. In particular, fine and intermediate scales exhibit substantial agreement with the predicted logarithmic covariance structure, while progressively coarser scales display increasing deviations associated with support-induced multiscale mixing.

The operational boundaries of the local freezing approximation are explicitly delineated by the scale-dependent evolution of both the reconstruction metrics and the empirical covariance bias. At the finest analysis scales, the~{\color{black} spatial} footprint of the analyzing wavelet remains sufficiently compact to isolate approximately homogeneous modulation neighbourhoods. In this highly localised regime, the multiplicative factorisation holds with exceptional fidelity; the energy-weighted relative error is tightly minimised, the Pearson correlation peaks, and the empirical covariance exhibits near-ideal logarithmic scaling. However, as the scale index progresses toward intermediate and coarse resolutions, the expanding wavelet support inevitably begins to encompass severe macroscopic variations within the underlying field. This progressive geometric breakdown of local freezing manifests clearly in the numerical data: mathematically as a rapid decline in reconstruction accuracy, and~{\color{black} spatial}ly as a structured covariance bias evidenced by the emergent resonance bump and residual mixing components. 

Crucially, the numerical results demonstrate that the overall resilience of this localised separation framework is fundamentally governed by the intermittency parameter, $\lambda^2$. As summarised in Table~\ref{tab: performance}, increasing the value of $\lambda^2$ substantially amplifies the structural dominance of the log-correlated modulation over the background carrier fluctuations. This heightened intermittency hardens the~{\color{black} spatial} footprint of the true modulation field, ensuring that the signal-to-noise ratio within the localised wavelet observation domain remains favourably skewed toward the multiplicative structural components. Consequently, for strongly intermittent fields, the high-energy fine-scale wavelet coefficients are able to lock onto the underlying modulation field with significantly greater accuracy. This structural dominance effectively suppresses the influence of carrier-induced noise and residual multiscale mixing, thereby extending the functional validity range of the freezing approximation and yielding markedly higher global reconstruction fidelity.

The support-controlled viewpoint also naturally explains why different wavelet families may produce different freezing behavior. Compact support, vanishing moments, regularity, and oscillatory structure all influence the effective probing geometry and therefore the quality of local modulation separation.

\subsection{Residual multiscale mixing and finite-scale breakdown}
One of the most important observations emerging from the numerical investigations is that residual covariance distortions are not merely implementation artifacts. Instead, many of the observed deviations appear to originate from finite-support multiscale mixing generated within the wavelet probing domain itself. This interpretation is particularly relevant for the covariance “bump” structures observed at intermediate and coarse scales. Initially, such deviations might be interpreted as reconstruction failure, numerical instability, imperfect surrogate correction, or limitations of the simulation procedure; however, the theoretical developments presented in Section~\ref{sec4} suggest a different interpretation. 

When the modulation field varies non-negligibly within the wavelet support, the local freezing approximation breaks down progressively. The wavelet coefficient then incorporates coupled contributions from multiple modulation neighbourhoods simultaneously, generating a residual covariance structure not captured by the idealised frozen approximation. From this perspective, the observed covariance distortions constitute experimentally observable signatures of residual multiscale mixing. This interpretation is particularly appealing because it transforms apparent methodological limitations into theoretically meaningful finite-scale effects. 

The numerical results suggest that fine scales remain dominated by localised freezing; intermediate scales exhibit partial support-induced coupling; and coarse scales enter regimes where modulation mixing dominates. This progressive transition appears consistently across multiple scales and realisations. To rigorously partition these physical mechanisms, surrogate kernel-subtraction experiments were employed to decouple the deterministic ~{\color{black}spatial} geometry of the analyzing wavelet from its dynamic interaction with the underlying modulation field. While the isolated white noise kernel successfully reproduces the localised resonance maximum at $r \approx 2^j$, confirming that the core peak is an intrinsic geometric artifact, it inherently fails to capture the full ~{\color{black}spatial} breadth of the empirical distortion. The subtraction of this baseline kernel isolates a highly structured residual mixing covariance, $C_{\mathrm{mix}}$, which systematically broadens the correlation footprint well beyond the wavelet's finite mathematical boundaries. Crucially, the persistence of this $C_{\mathrm{mix}}$ component provides direct numerical evidence for an effective support extension, demonstrating that the long-range ~{\color{black}spatial} memory of the log-correlated background field dynamically couples with the wavelet to generate emergent interaction scales that cannot be explained by the isolated filter geometry alone.

The possibility that effective support geometry itself evolves through modulation-induced coupling is especially intriguing. If confirmed, this would suggest that localised multiscale operators may develop emergent interaction scales exceeding nominal wavelet support widths. Such effects may potentially connect localised modulation extraction to broader questions involving multiscale operator renormalisation, support-dependent stochastic coupling, and localised regularity transitions. At present, these interpretations remain preliminary and require additional theoretical investigation.

\subsection{Relation to wavelet regularity theory and multifractal analysis}
The localised freezing framework developed here naturally connects to wavelet regularity theory and singularity analysis. Wavelet coefficients are known to encode local Hölder regularity through scale-dependent decay laws of the form
\begin{equation}
\lvert W_X(a,b) \rvert \sim a^{h(b)},
\end{equation}
where $h(b)$ denotes the local regularity at position $b$. Foundational work by Jaffard and collaborators established deep connections between wavelet coefficients, local singularity structure, and multifractal geometry. The present work suggests that local modulation freezing may admit a complementary interpretation within this regularity framework.

The validity of localised multiplicative separation depends directly on how rapidly the modulation field varies across the wavelet support. Local smoothness of the modulation process therefore controls the quality of freezing. Regions possessing sufficiently regular local organisation naturally favor approximate modulation constancy within the probing domain, whereas strongly heterogeneous or rapidly varying regions generate enhanced multiscale coupling. From this perspective, freezing validity itself becomes a local regularity phenomenon. Importantly, this interpretation differs conceptually from classical WTMM multifractal analysis. In WTMM frameworks, localisation primarily serves to identify singular structures whose statistics are subsequently aggregated globally in order to estimate multifractal spectra. In the present work, localisation instead acts directly as the operational mechanism underlying local modulation extraction. The distinction is subtle but fundamental.

Historically, locality constituted an important conceptual foundation of wavelet singularity analysis. However, the localised wavelet coefficient itself was rarely interpreted as a finite-support operator capable of inducing local multiplicative separation. The present framework attempts to develop precisely this viewpoint, potentially providing new connections between localised intermittency analysis, inverse multifractal problems, local regularity estimation, and stochastic modulation theory.

\section{Conclusions, broader implications and future directions}\label{Conclusions}
The present work was motivated primarily by the problem of localised MRW unwrapping. However, the framework developed here may admit substantially broader applicability. Because the proposed freezing mechanism depends primarily on localisation and support geometry rather than on a specific modulation ontology, the formulation naturally extends beyond classical MRW models toward more general multiplicatively modulated multiscale systems.

Although the present work focused mainly on one-dimensional formulations for analytical clarity, the localised freezing framework extends naturally to higher-dimensional intermittent fields, where support geometry, overlap structure, and multiscale coupling become directly spatially observable. From this perspective, localisation itself acquires an operational geometric meaning, linking wavelet support structure to the local organisation of multiscale stochastic interactions.

Potential extensions of the framework include nonstationary intermittency, localised multifractal analysis, stochastic texture modelling, multiscale inverse problems, adaptive regularity estimation, and more general localised stochastic operator formulations. The present formulation may also provide a basis for revisiting earlier localised multifractal approaches from a more rigorous operator-theoretic perspective. In particular, the explicit interpretation of support-controlled freezing and residual multiscale mixing may help clarify the relation between local singularity organisation and multiscale statistical coupling.

Several important theoretical questions nevertheless remain open. A precise mathematical characterisation of freezing validity as a function of support geometry, scale, and local regularity requires further investigation. Similarly, the structure of residual multiscale mixing kernels remains only partially understood, especially in regimes where effective support overlap becomes increasingly nonlocal. The relation between localised modulation extraction and broader wavelet operator theory likewise deserves substantially deeper analysis.

Additional work will also be required to characterise the dependence of the framework on wavelet family selection, dimensionality, anisotropic modulation structure, non-Gaussian modulation fields, and adaptive support-selection strategies. These extensions may prove particularly important in higher-dimensional settings where support geometry itself becomes dynamically structured across scales.

Nevertheless, the results presented here strongly suggest that localised wavelet operators may provide direct access to local multiplicative organisation itself rather than merely to its global statistical signatures. Within this framework, wavelet localisation no longer acts solely as a passive decomposition principle. Instead, localised freezing transforms localisation into an active multiscale probing mechanism whose validity depends on support geometry, finite-scale overlap, and local modulation variability.

From this viewpoint, localisation becomes not only a representational property of wavelet transforms, but a fundamental operational mechanism for probing multiscale stochastic structure.

\section*{Author Contributions}
M.P. performed the numerical simulations, implemented the wavelet-based reconstruction framework, generated the figures, and contributed to the analytical development and manuscript preparation. Z.R.S. conceived and developed the theoretical framework, including the localised multiplicative modulation interpretation, freezing formalism, and regularity-based analysis. Both authors contributed to the interpretation of the results, development of the manuscript structure, and writing of the manuscript.

\section*{Data Availability}
Code and simulation datasets are available on reasonable request through the corresponding author.

\nocite{*}
\bibliography{Bibliography}

@article{Lakhal25,
    author    = {Lakhal, S. and Ponson, L. and Benzaquen, M. and Bouchaud, J. P.},
    title     = {Wrapping and unwrapping multifractal fields: Application to fatigue and abrupt failure fracture surfaces},
    journal   = {Phys. Rev. Research},
    volume    = {7},
    number    = {1},
    pages     = {L012003},
    year      = {2025}
}

@article{Kang14,
    author    = {Kang, Peter K. and de Anna, Pietro and Nunes, Joao P. and Bijeljic, Branko and Blunt, Martin J. and Juanes, Ruben},
    title     = {Pore-scale intermittent velocity structure underpinning anomalous transport through 3-D porous media},
    journal   = {Geophysical Research Letters},
    volume    = {41},
    number    = {17},
    pages     = {6184--6190},
    doi       = {10.1002/2014GL061475},
    year      = {2014}
}

@article{Muzy06,
    author    = {Muzy, J. F. and Bacry, E. and Kozhemyak, A.},
    title     = {Extreme values and fat tails of multifractal fluctuations},
    journal   = {Phys. Rev. E},
    volume    = {73},
    number    = {6},
    pages     = {066114},
    numpages  = {14},
    year      = {2006},
    publisher = {American Physical Society},
    doi       = {10.1103/PhysRevE.73.066114}
}

@article{SerGiacomi15,
    author    = {Ser-Giacomi, Enrico and Rossi, Vincent and López, Cristóbal and Hernández-García, Emilio},
    title     = {Flow networks: A characterization of geophysical fluid transport},
    journal   = {Chaos: An Interdisciplinary Journal of Nonlinear Science},
    volume    = {25},
    number    = {3},
    pages     = {036404},
    year      = {2015},
    doi       = {10.1063/1.4908231}
}

@article{Schmitt99,
    author    = {Schmitt, François and Schertzer, Daniel and Lovejoy, Shaun},
    title     = {Multifractal analysis of foreign exchange data},
    journal   = {Applied Stochastic Models and Data Analysis},
    volume    = {15},
    number    = {1},
    pages     = {29--53},
    year      = {1999}
}

@article{SuarezGarcia14,
    author    = {Suárez-García, P. and Gómez-Ullate, D.},
    title     = {Multifractality and long memory of a financial index},
    journal   = {Physica A: Statistical Mechanics and its Applications},
    volume    = {394},
    pages     = {226--234},
    year      = {2014},
    issn      = {0378-4371},
    doi       = {10.1016/j.physa.2013.09.038}
}

@article{Friedrich11,
    author    = {Friedrich, R. and Peinke, J. and Sahimi, M. and Reza Rahimi Tabar, M.},
    title     = {Approaching complexity by stochastic methods: From biological systems to turbulence},
    journal   = {Physics Reports},
    volume    = {506},
    number    = {5},
    pages     = {87--162},
    year      = {2011},
    issn      = {0370-1573},
    doi       = {10.1016/j.physrep.2011.05.003}
}

@article{Granero24,
    author    = {Granero-Belinchon, Carlos and Roux, Stephane G. and Garnier, Nicolas},
    title     = {Multiscale and Anisotropic Characterization of Images Based on Complexity: an Application to Turbulence},
    journal   = {Physica D: Nonlinear Phenomena},
    volume    = {459},
    pages     = {134027},
    year      = {2024},
    publisher = {Elsevier},
    doi       = {10.1016/j.physd.2023.134027}
}

@article{daubechies88,
    author    = {Daubechies, Ingrid},
    title     = {Orthonormal bases of compactly supported wavelets},
    journal   = {Communications on Pure and Applied Mathematics},
    volume    = {41},
    number    = {7},
    pages     = {909--996},
    year      = {1988},
    publisher = {Wiley},
    doi       = {10.1002/cpa.3160410705}
}

@book{daubechies92,
    author    = {Daubechies, Ingrid},
    title     = {Ten Lectures on Wavelets},
    series    = {CBMS-NSF Regional Conference Series in Applied Mathematics},
    volume    = {61},
    year      = {1992},
    publisher = {Society for Industrial and Applied Mathematics (SIAM)},
    address   = {Philadelphia, PA},
    isbn      = {978-0-89871-274-2},
    doi       = {10.1137/1.9781611970104}
}

@article{Jimenez2000,
    author    = {Jiménez, Javier},
    title     = {Intermittency and cascades},
    journal   = {Journal of Fluid Mechanics},
    volume    = {409},
    pages     = {99--120},
    year      = {2000},
    doi       = {10.1017/s0022112099007739}
}

@article{Borlandetal2005,
    author    = {Borland, Lisa and Bouchaud, Jean-Philippe and Muzy, Jean-Francois and Zumbach, Gilles},
    title     = {The Dynamics of Financial Markets – Mandelbrot's multifractal cascades, and beyond},
    journal   = {arXiv},
    year      = {2005},
    doi       = {10.48550/arxiv.cond-mat/0501292}
}

@article{Mandelbrot1974,
    author    = {Mandelbrot, Benoit B.},
    title     = {Intermittent turbulence in self-similar cascades: divergence of high moments and dimension of the carrier},
    journal   = {Journal of Fluid Mechanics},
    volume    = {62},
    pages     = {331--358},
    year      = {1974},
    doi       = {10.1017/s0022112074000711}
}

@article{Bacryetal2000,
    author    = {Bacry, E. and Delour, J. and Muzy, J. F.},
    title     = {A multifractal random walk},
    journal   = {arXiv},
    year      = {2000},
    doi       = {10.1103/PhysRevE.64.026103}
}

@article{RobertVargas2010,
    author    = {Robert, Raoul and Vargas, Vincent},
    title     = {Gaussian multiplicative chaos revisited},
    journal   = {The Annals of Probability},
    volume    = {38},
    year      = {2010},
    doi       = {10.1214/09-aop490}
}

@article{Abryetal2009,
    author    = {Abry, Patrice and Chainais, Pierre and Coutin, Laure and Pipiras, Vladas},
    title     = {Multifractal Random Walks as Fractional Wiener Integrals},
    journal   = {IEEE Transactions on Information Theory},
    volume    = {55},
    pages     = {3825--3846},
    year      = {2009},
    doi       = {10.1109/tit.2009.2023708}
}

@article{RhodesVargas2014,
    author    = {Rhodes, Rémi and Vargas, Vincent},
    title     = {Gaussian multiplicative chaos and applications: A review},
    journal   = {Probability Surveys},
    volume    = {11},
    year      = {2014},
    doi       = {10.1214/13-ps218}
}

@article{MallatHwang1992,
    author    = {Mallat, Stéphane and Hwang, Wen Liang},
    title     = {Singularity detection and processing with wavelets},
    journal   = {IEEE Transactions on Information Theory},
    volume    = {38},
    number    = {2},
    pages     = {617--643},
    year      = {1992},
    doi       = {10.1109/18.119727}
}

@article{Muzyetal1991,
    author    = {Muzy, J. F. and Bacry, E. and Arneodo, A.},
    title     = {Wavelets and multifractal formalism for singular signals: Application to turbulence data},
    journal   = {Physical Review Letters},
    volume    = {67},
    number    = {25},
    pages     = {3515--3518},
    year      = {1991},
    doi       = {10.1103/PhysRevLett.67.3515}
}

@article{Muzyetal1993,
    author    = {Muzy, J. F. and Bacry, E. and Arneodo, A.},
    title     = {Multifractal formalism for fractal signals: The structure-function approach versus the wavelet-transform modulus-maxima method},
    journal   = {Physical Review E},
    volume    = {47},
    number    = {2},
    pages     = {875--884},
    year      = {1993},
    doi       = {10.1103/PhysRevE.47.875}
}

@article{Jaffard1997a,
    author    = {Jaffard, Stéphane},
    title     = {Multifractal Formalism for Functions Part I: Results Valid For All Functions},
    journal   = {SIAM Journal on Mathematical Analysis},
    volume    = {28},
    number    = {4},
    pages     = {944--970},
    year      = {1997},
    doi       = {10.1137/S0036141095282991}
}

@article{Jaffard1997b,
    author    = {Jaffard, Stéphane},
    title     = {Multifractal Formalism for Functions Part II: Self-Similar Functions},
    journal   = {SIAM Journal on Mathematical Analysis},
    volume    = {28},
    number    = {4},
    pages     = {971--998},
    year      = {1997},
    doi       = {10.1137/S0036141095283005}
}

@inproceedings{FrischParisi1985,
    author    = {Frisch, Uriel and Parisi, Giorgio},
    title     = {Fully Developed Turbulence and Intermittency},
    booktitle = {Turbulence and Predictability in Geophysical Fluid Dynamics and Climate Dynamics},
    editor    = {Ghil, M. and Benzi, R. and Parisi, G.},
    publisher = {North-Holland},
    address   = {Amsterdam},
    pages     = {84--88},
    year      = {1985}
}

@article{Wendtetal2007,
    author    = {Wendt, Herwig and Abry, Patrice and Jaffard, Stéphane},
    title     = {Bootstrap for empirical multifractal analysis},
    journal   = {IEEE Signal Processing Magazine},
    volume    = {24},
    number    = {4},
    pages     = {38--48},
    year      = {2007},
    doi       = {10.1109/MSP.2007.4286563}
}

@article{Lashermesetal2004,
    author    = {Lashermes, Bruno and Jaffard, Stéphane and Abry, Patrice},
    title     = {Wavelet leader based multifractal analysis},
    journal   = {IEEE International Conference on Acoustics, Speech, and Signal Processing},
    volume    = {4},
    pages     = {161--164},
    year      = {2004},
    doi       = {10.1109/ICASSP.2004.1326781}
}

@book{Mallat1999,
  author    = {Mallat, Stéphane},
  title     = {A Wavelet Tour of Signal Processing},
  edition   = {2nd},
  publisher = {Academic Press},
  year      = {1999},
  isbn      = {978-0124666061},
  doi       = {10.1016/B978-012466606-1/50008-8}
}

\end{document}